\documentclass{article}

\usepackage{arxiv}
\usepackage[utf8]{inputenc} 
\usepackage[T1]{fontenc}    
\usepackage{hyperref}       
\usepackage{url}            
\usepackage{booktabs}       
\usepackage{amsfonts}       
\usepackage{nicefrac}       
\usepackage{microtype}      
\usepackage{lipsum}
\usepackage{amsmath}
\usepackage{amssymb} 
\usepackage{graphicx}
\usepackage{mathrsfs}
\usepackage{tikz}
\usepackage{cite}
\usepackage{color}
\usetikzlibrary{arrows.meta}

\title{Electron in bilayer graphene with magnetic fields leading to shape invariant potentials}
\author{
David J Fern{\'a}ndez C, Juan D Garc{\'i}a M and Daniel O-Campa\\
Physics Department, Cinvestav\\
P.O.B. 14-740, 07000 Mexico City, Mexico\\
e-mail: david@fis.cinvestav.mx, dgarcia@fis.cinvestav.mx, dortiz@fis.cinvestav.mx
}

\begin{document}
\maketitle

\begin{abstract}
The quantum behavior of electrons in bilayer graphene with applied magnetic fields is addressed. By using second-order supersymmetric quantum mechanics the problem is transformed into two intertwined one dimensional stationary Schr{\"o}dinger equations whose potentials are required to be shape invariant. Analytical solutions for the energy bound states are obtained for several magnetic fields. The associated spectrum is analyzed, and the probability and current densities are determined.
\end{abstract}

\section{Introduction}

Graphene is the thinnest material ever known, which is composed of a single layer of carbon atoms arranged in a hexagonal lattice. It was experimentally isolated in 2004 by Geim and Novoselov, and due to its outstanding electric properties plenty of studies have been performed, one of the most important being the integer quantum Hall effect. Also, it is known that the charge carriers in monolayer graphene behave as massless chiral quasiparticles with a linear dispersion relation, leading to a description in terms of a Dirac-like effective Hamiltonian \cite{1}.

On the other hand, although many properties of bilayer graphene are similar to the monolayer ones, for bilayer graphene at low energies the integer quantum Hall effect indicates the presence of massive chiral quasiparticles with a parabolic dispersion relation instead of a linear one. Moreover, bilayer graphene also can have applications in electronic devices, as well as in many other areas of technology \cite{1}.

In this article we are going to consider the interaction of electrons in bilayer graphene with magnetic fields orthogonal to the layer surfaces, which are invariant under translations along a fixed given direction on the plane. It is worth to notice that this physical configuration is similar to the one usually addressed for monolayer graphene. In fact, plenty of exact solutions for monolayer graphene in orthogonal magnetic fields have been recently found by using first-order supersymmetric quantum mechanic \cite{10,mtm11,7,df17,chrv18,llr18,cf20}. In this article we are going to show that for bilayer graphene placed in magnetic fields similar to those addressed in \cite{10} exact solutions for the effective Hamiltonian can be found by using second-order SUSY QM \cite{ais93,aicd95,sa99,6,fe10,fe19}.

This paper is organized as follows: in section 2 we will introduce the effective Hamiltonian ruling the bilayer graphene in external magnetic fields and the SUSY QM approach useful to address the problem. Section 3 contains several kinds of magnetic fields that can be solved exactly through this method, and a discussion of these results. Our conclusions are contained in section 4.
\section{Effective Hamiltonian for bilayer graphene}

\begin{figure}[ht]
\centering
\begin{tikzpicture}[xscale=0.94,yscale=0.72] 


\foreach \n in {0,1,2,3}
{
\ifodd \n \newcommand{\xo}{3/2*\n} \newcommand{\yo}{-0.866}
\else \newcommand{\xo}{3/2*\n} \newcommand{\yo}{0} 
\fi    
\foreach \c in {0,1,2,3}
{
\foreach \x in {0,1,3,4} 
{
\foreach \y in {0,1,-1} 
{
\ifnum \x=0 \ifnum \y=0 \fill[color=black] (\yo+\y*0.866+\c*1.732,\xo+\x/2) circle[radius=3pt]; 
\draw (\yo+\y*0.866-0.866+\c*1.732,\xo+\x/2+1/2)--(\yo+\y*0.866+\c*1.732,\xo+\x/2)--(\yo+\y*0.866+0.866+\c*1.732,\xo+\x/2+1/2); \fi \fi
 \ifnum \x=1 \ifnum \y=1 \draw (\yo+\y*0.866+\c*1.732,\xo+\x/2)--(\yo+\y*0.866+\c*1.732,\xo+\x/2+1); \fill[color=gray] (\yo+\y*0.866+\c*1.732,\xo+\x/2) circle [radius=3pt]; 
 \fi
 \ifnum \y=-1 \draw (\yo+\y*0.866+\c*1.732,\xo+\x/2)--(\yo+\y*0.866+\c*1.732,\xo+\x/2+1); \fill[color=gray] (\yo+\y*0.866+\c*1.732,\xo+\x/2) circle [radius=3pt]; 
 \fi \fi
 \ifnum \x=3 \ifnum \y=1 \draw (\yo+\y*0.866+\c*1.732,\xo+\x/2)--(\yo+\y*0.866-0.866+\c*1.732,\xo+\x/2+1/2); \fill[color=black] (\yo+\y*0.866+\c*1.732,\xo+\x/2) circle [radius=3pt]; \fi
 \ifnum \y=-1 \draw (\yo+\y*0.866+\c*1.732,\xo+\x/2)--(\yo+\y*0.866+0.866+\c*1.732,\xo+\x/2+1/2); \fill[color=black] (\yo+\y*0.866+\c*1.732,\xo+\x/2) circle [radius=3pt]; \fi \fi
 \ifnum \x=4 \ifnum \y=0 \fill[color=gray] (\yo+\y*0.866+\c*1.732,\xo+\x/2) circle [radius=3pt]; 
 \fi \fi
}
}
}
}
\draw[arrows={|Stealth[scale=1.5]-Stealth[scale=1.5]|}] (0,-0.3)-- node[below=1pt]{\Large $a$} (1.732,-0.3); 
\draw[arrows={-Stealth[scale=1]},line width=2pt] (1.732,1)--(2.598,2.5) node[right=0.5pt]{\Large $\vec{a}_1$};
\draw[arrows={-Stealth[scale=1]},line width=2pt] (1.732,1)--(0.866,2.5) node[left=0.5pt]{\Large $\vec{a}_2$};
\draw[dashed,line width=2pt] (2.598,2.5)--(1.732,4) (0.866,2.5)--(1.732,4);
\draw (1.732,2) node[below=1pt]{$B$};
\draw (1.732,3) node[above=1pt]{$A$};
\end{tikzpicture}
\newcommand{\w}{1.75}
\newcommand{\h}{0.4}
\newcommand{\nx}{0.5}
\newcommand{\ny}{0.8}
\newcommand{\cx}{-0.25}
\newcommand{\cy}{2.5}
\begin{tikzpicture}  


\foreach \s in {1,2} 
{
\ifnum \s=1
 \foreach \t in {0,1,2,3} 
 {
 \filldraw[fill=white] (\w*\t,\h*\t)--(\w*\t+1,\h*\t)--(\w*\t+1.75,\h*\t+0.4)--(\w*\t+1.5,\h*\t+0.8)--(\w*\t+0.5,\h*\t+0.8)--(\w*\t-0.25,\h*\t+0.4)--(\w*\t,\h*\t);
 \foreach \y in {0,4,8}
 {
 \foreach \x in {0,1,5,15,175,-25}
 {
 \ifnum \y=0 \ifnum \x=0 \fill[lightgray] (\w*\t+\x,\h*\t+\y) circle [radius=3pt]; \fi \ifnum \x=1 \fill[black] (\w*\t+\x,\h*\t+\y) circle [radius=3pt]; \fi \fi
 \ifnum \y=4 \ifnum \x=175 \fill[lightgray] (\w*\t+\x/100,\h*\t+\y/10) circle [radius=3pt]; \fi \ifnum \x=-25 \fill[black] (\w*\t+\x/100,\h*\t+\y/10) circle [radius=3pt]; \fi \fi
 \ifnum \y=8 \ifnum \x=5 \fill[lightgray] (\w*\t+\x/10,\h*\t+\y/10) circle [radius=3pt]; \fi \ifnum \x=15 \fill[black] (\w*\t+\x/10,\h*\t+\y/10) circle [radius=3pt]; \fi \fi
 }
 }
 }
 \else
 \foreach \t in {0,1,2}
 {
 \filldraw[fill=white] (\w*\t+\nx,\h*\t+\ny)--(\w*\t+\nx-0.25,\h*\t+\ny+0.4)--(\w*\t+\nx+0.5,\h*\t+\ny+0.8)--(\w*\t+\nx+1.5,\h*\t+\ny+0.8)--(\w*\t+\nx+1.75,\h*\t+\ny+0.4)--(\w*\t+\nx+1,\h*\t+\ny)--(\w*\t+\nx,\h*\t+\ny);
 \foreach \y in {0,4,8}
 {
 \foreach \x in {0,1,5,15,175,-25}
 {
 \ifnum \y=0 \ifnum \x=0 \fill[lightgray] (\w*\t+\x+\nx,\h*\t+\y+\ny) circle [radius=3pt]; \fi \ifnum \x=1 \fill[black] (\w*\t+\x+\nx,\h*\t+\y+\ny) circle [radius=3pt]; \fi \fi
 \ifnum \y=4 \ifnum \x=175 \fill[lightgray] (\w*\t+\nx+\x/100,\h*\t+\ny+\y/10) circle [radius=3pt]; \fi \ifnum \x=-25 \fill[black] (\w*\t+\nx+\x/100,\h*\t+\ny+\y/10) circle [radius=3pt]; \fi \fi
 \ifnum \y=8 \ifnum \x=5 \fill[lightgray] (\w*\t+\nx+\x/10,\h*\t+\ny+\y/10) circle [radius=3pt]; \fi \ifnum \x=15 \fill[black] (\w*\t+\nx+\x/10,\h*\t+\ny+\y/10) circle [radius=3pt]; \fi \fi
 }
 }
 }
 \fi
}

\foreach \s in {1,2} 
{
\ifnum \s=1
 \foreach \t in {0,1,2,3} 
 {
 \filldraw[fill=white] (\w*\t+\cx,\h*\t+\cy)--(\w*\t+\cx+1,\h*\t+\cy)--(\w*\t+\cx+1.75,\h*\t+\cy+0.4)--(\w*\t+\cx+1.5,\h*\t+\cy+0.8)--(\w*\t+\cx+0.5,\h*\t+\cy+0.8)--(\w*\t+\cx-0.25,\h*\t+\cy+0.4)--(\w*\t+\cx,\h*\t+\cy);
 \foreach \y in {0,4,8}
 {
 \foreach \x in {0,1,5,15,175,-25}
 {
 \ifnum \y=0 \ifnum \x=0 \fill[black] (\w*\t+\x+\cx,\h*\t+\y+\cy) circle [radius=3pt]; \fi \ifnum \x=1 \fill[gray] (\w*\t+\x+\cx,\h*\t+\y+\cy) circle [radius=3pt]; \fi \fi
 \ifnum \y=4 \ifnum \x=175 \fill[black] (\w*\t+\cx+\x/100,\h*\t+\cy+\y/10) circle [radius=3pt]; \fi \ifnum \x=-25 \fill[gray] (\w*\t+\cx+\x/100,\h*\t+\cy+\y/10) circle [radius=3pt]; \fi \fi
 \ifnum \y=8 \ifnum \x=5 \fill[black] (\w*\t+\cx+\x/10,\h*\t+\cy+\y/10) circle [radius=3pt]; \fi \ifnum \x=15 \fill[gray] (\w*\t+\cx+\x/10,\h*\t+\cy+\y/10) circle [radius=3pt]; \fi \fi
 }
 }
 }
 \else
 \foreach \t in {0,1,2}
 {
 \filldraw[fill=white] (\w*\t+\cx+\nx,\h*\t+\cy+\ny)--(\w*\t\cx+\nx-0.25,\h*\t+\cy+\ny+0.4)--(\w*\t+\cx+\nx+0.5,\h*\t+\cy+\ny+0.8)--(\w*\t+\cx+\nx+1.5,\h*\t+\cy+\ny+0.8)--(\w*\t+\cx+\nx+1.75,\h*\t+\cy+\ny+0.4)--(\w*\t+\cx+\nx+1,\h*\t+\cy+\ny)--(\w*\t+\cx+\nx,\h*\t+\cy+\ny);
 \foreach \y in {0,4,8}
 {
 \foreach \x in {0,1,5,15,175,-25}
 {
 \ifnum \y=0 \ifnum \x=0 \fill[black] (\w*\t+\cx+\x+\nx,\h*\t+\cy+\y+\ny) circle [radius=3pt]; \fi \ifnum \x=1 \fill[gray] (\w*\t+\cx+\x+\nx,\h*\t+\cy+\y+\ny) circle [radius=3pt]; \fi \fi
 \ifnum \y=4 \ifnum \x=175 \fill[black] (\w*\t+\cx+\nx+\x/100,\h*\t+\cy+\ny+\y/10) circle [radius=3pt]; \fi \ifnum \x=-25 \fill[gray] (\w*\t+\cx+\nx+\x/100,\h*\t+\cy+\ny+\y/10) circle [radius=3pt]; \fi \fi
 \ifnum \y=8 \ifnum \x=5 \fill[black] (\w*\t\cx+\nx+\x/10,\h*\t+\cy+\ny+\y/10) circle [radius=3pt]; \fi \ifnum \x=15 \fill[gray] (\w*\t+\cx+\nx+\x/10,\h*\t+\cy+\ny+\y/10) circle [radius=3pt]; \fi \fi
 }
 }
 }
 \fi
}

\draw[thick] (-0.25,0.5)--(-0.25,2.4)  (3.25,1.3)--(3.25,3.2) (5,1.7)--(5,3.6) (6.75,2.1)--(6.75,4);
\draw (0.25,1.3)--(0.25,2.5) (2,1.7)--(2,2.9) (3.75,2.1)--(3.75,3.3) (5.5,2.5)--(5.5,3.7);
\draw[dashed] (0.25,2.5)--(0.25,3.2) (2,2.9)--(2,3.6) (3.75,3.3)--(3.75,4) (5.5,3.7)--(5.5,4.4);
\draw[ultra thin] (1.5,0.9)--(1.5,2.8);
\draw[ultra thin, dashed] (0.5,0.89)--(1.5,2.9);
\draw[ultra thin, dashed] (3.5,0.89)--(4.25,3.25);

\fill[black] (1.75,0.4) node[anchor=north]{$B1$};
\fill[black] (2.75,0.4) node[anchor=north]{$A1$};
\fill[black] (0.75,4.1) node[anchor=south]{$A2$};
\fill[black] (1.75,4.1) node[anchor=south]{$B2$};
\draw (-0.25,1.5) node[anchor=east]{\large $\gamma_1$};
\draw (0.5,0) node[anchor=north]{\large $\gamma_0$}; 
\draw (3,4.5) node[anchor=south]{\large $\gamma_0$};
\draw (1,2) node[anchor=east]{\large $\gamma_4$};
\draw (4.7,2.8) node[anchor=east]{\large $\gamma_3$};
\draw (6,1.6) node[anchor=east]{\large $a$};

\draw[<->, thick] (5.7,1.2)--(6.3,2);
\end{tikzpicture}
\caption{The structure of monolayer graphene, where atoms A (B) are shown as black (gray) circles, $\vec{a}_1$ and $\vec{a}_2$ are the primitive lattice vectors and the rhombus is the conventional unit cell (left). Side view of the bilayer graphene, where atoms A1, B1 on the lower layer are shown as black and light gray circles while atoms A2, B2 on the upper layer are drawn as black and gray circles, respectively (right).} \label{Imagen2.1}
\end{figure}
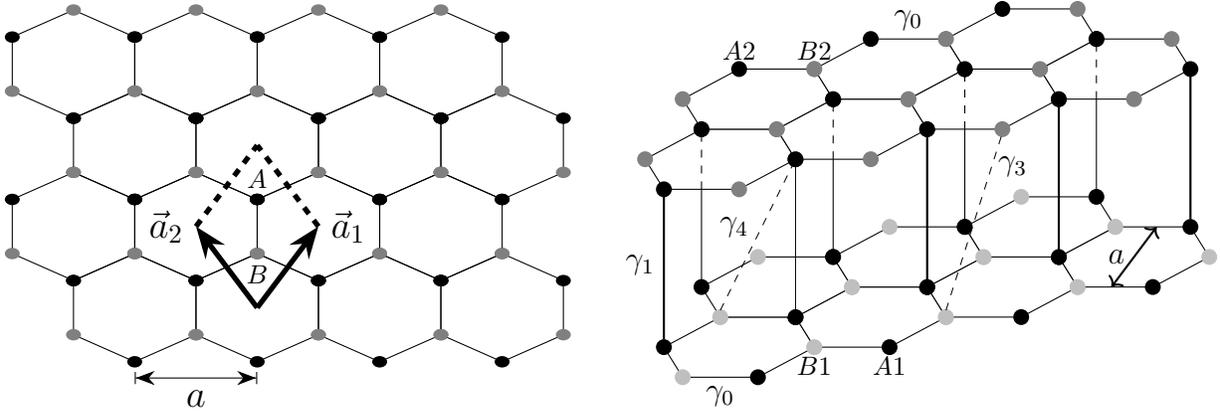

\begin{figure}[ht]
\centering
\includegraphics[scale=0.7]{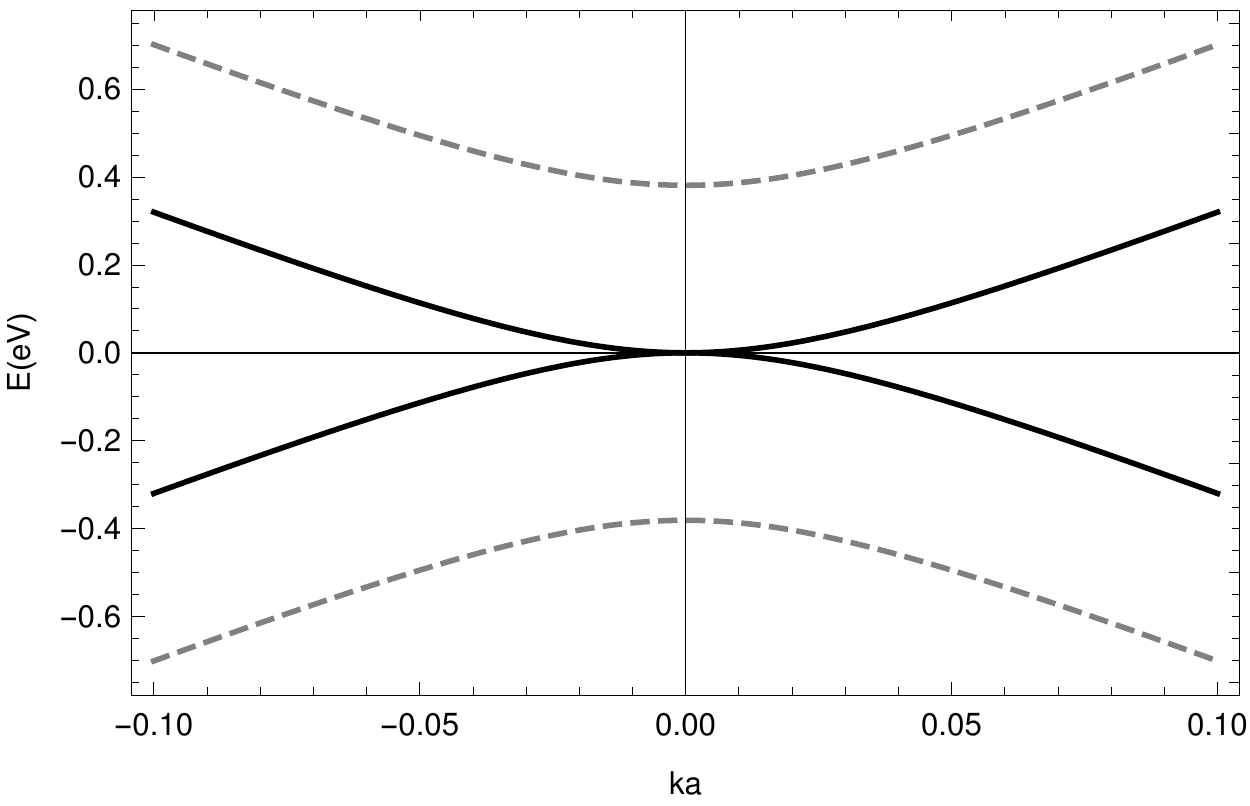}
\caption{The band structure of bilayer graphene within the tight-binding model.} \label{Imagen2.2}
\end{figure}

Bilayer graphene is a material composed by two monolayers of carbon atoms, each one having a honeycomb crystal structure. Its electronic properties can be studied inside the so-called tight-binding model\cite{1,2,3,4}. Figures \ref{Imagen2.1} (a and b) illustrate the structure of monolayer and bilayer graphene, respectively. We can notice that the second layer is rotated by an angle of $\frac{\pi}{3}$ with respect to the first layer; this is known as Bernal stacking in the literature and it is the most common form of bilayer graphene appearing in nature\cite{1}. Also, the sublattices A of each layer lie exactly on top of one another, with a hopping parameter $\gamma_1$ between them, whereas there are no hopping processes between the up and down sublattices B. The parameter $\gamma_1$ is usually taken as $\gamma_1=0.4$ eV, which is one order of magnitude lower than the nearest-neighbour in-plane hopping parameter $\gamma_0\approx 2.97$ eV. This simple model is described by the Hamiltonian
\begin{equation}
H(\vec{k})=\left(
\begin{array}{cccc} 
0 & \gamma_0S(\vec{k}) & \gamma_1 & 0 \\
\gamma_0 S^{*}(\vec{k}) & 0 & 0 & 0 \\
\gamma_1 & 0 & 0 & \gamma_0 S^{*}(\vec{k}) \\
0 & 0 & \gamma_0S(\vec{k}) & 0
\end{array} \right),
\label{2.1} 
\end{equation}
where $S(\vec{k})$ is given by
\begin{equation}
S(\vec{k})=2\exp\left(\frac{ik_{x}a}{2\sqrt{3}}\right)\cos\left(\frac{k_{y}a}{2}\right)+\exp\left(-\frac{ik_{x}a}{\sqrt{3}}\right).
\label{2.2}
\end{equation}
The ordering of the basis states is: first layer sublattice A, first layer sublattice B, second layer sublattice A and
second layer sublattice B. The matrix in Eq.~(\ref{2.1}) can be easily diagonalized, with its eigenvalues being given by
\begin{equation}
E(\vec{k})=\pm\frac{1}{2}\gamma_1\pm\sqrt{\frac{1}{4}\gamma_1^2+\gamma_0^2 |S(\vec{k})|^{2}},
\label{2.3}
\end{equation}
with two independent $\pm$ signs; this spectrum is shown in Figure \ref{Imagen2.2}. As can be seen, two bands touch to each other at the points $K$ and $K^{'}$. Around these points the corresponding eigenvalues take the following form
\begin{equation}
E(\vec{k})_{1,2}\approx\pm\frac{\gamma_0^2|S(\vec{k})|^{2}}{\gamma_1}\approx\pm\frac{\hbar^{2} q^{2}}{2 m^{*}},
\label{2.4}
\end{equation}
where $\vec{q}$ could be either $\vec{k}-\vec{K}$ or $\vec{k}-\vec{K}^{'}$, and $m^{*}=\frac{\gamma_1}{2{v_F}^{2}}\approx0.054m_e$ is the electron effective mass, with $m_e$ being the electron free mass. This tells us that bilayer graphene has a parabolic band structure without an energy gap. The other two branches $E(\vec{k})_{3,4}$ have a gap of size $2\gamma_1$, thus they can be neglected at low-energies. Taking into account that atoms $A1$ and $A2$ are in dimer sites, which are coupled by a strong interlayer coupling \cite{5}, and that $\hbar q_x$ and $\hbar q_y$ are the operators $p_x=-i\hbar\frac{\partial}{\partial x}$ and $p_y=-i\hbar\frac{\partial}{\partial y}$, the following effective Hamiltonian around $\vec{K}$ is obtained
\begin{equation}
H_{\vec{K}}=\frac{1}{2m^{*}}\left(
\begin{array}{cc} 
0 &\left(p_x-ip_y\right)^{2}\\
\left(p_x+ip_y\right)^{2} & 0
\end{array} \right),
\label{2.5}
\end{equation}
which is neither a Dirac-like (relativistic case) nor a Schrödinger-like (non-relativistic case) effective Hamiltonian. However, it has very special properties like the chirality of its eigenstates and also that $H_{\vec{K}^{'}}=H_{\vec{K}}^{T}$; in this work we are going to study just $H_{\vec{K}}$, and by simplicity we will denote it as $H$\cite{1,5}.

If a magnetic field is applied, the minimal coupling rule transforms $p_i$ into $p_i+\frac{e}{c}A_i$ in Eq.~(\ref{2.5}). We are going to consider magnetic fields which are orthogonal to the graphene surface (plane ($x-y$)) and change only along a fixed direction ($x$). Thus, in the Landau gauge the vector potential can be chosen as $\vec{A}=\mathcal{A}(x)\hat{e}_y$, which implies that $\vec{B}=\mathcal{B}(x)\hat{e}_z$, $\mathcal{B}(x)=\mathcal{A}'(x)$. Now, the eigenvalue equation defined by the Hamiltonian (\ref{2.5}) reads
\begin{equation}
H\Psi\left(x,y\right)=\frac{1}{2m^{*}}\left(
\begin{array}{cc} 
0 & \Pi^2\\
\left(\Pi^{\dagger}\right)^{2} & 0
\end{array} \right)
\Psi\left(x,y\right) =E\Psi\left(x,y\right),
\label{2.6}
\end{equation} 
where $\Pi=p_x-ip_y-i\frac{e}{c}\mathcal{A}(x)$. Taking into account the invariance of this equation under translations along $y$-direction, it is natural to propose the following form for $\Psi\left(x,y\right)$: 
\begin{equation}
\Psi\left(x,y\right)=e^{iky}\left(
\begin{array}{c} 
\psi^{\left(2\right)}\left(x\right)\\ 
\psi^{\left(0\right)}\left(x\right)
\end{array}\right), 
\label{2.7}
\end{equation}
with $k$ being the wave-number in $y$-direction. By plugging Eq.~(\ref{2.7}) into Eq.~(\ref{2.6}) it is obtained the following system of equations
\begin{align}
L^{-}_2\psi^{\left(0\right)}\left(x\right)&=
-\mathscr{E}\psi^{\left(2\right)}\left(x\right),\label{2.8} \\
L^{+}_2\psi^{\left(2\right)}\left(x\right)&=
-\mathscr{E}\psi^{\left(0\right)}\left(x\right),\label{2.9}
\end{align}
where $L^{-}_2$ and $L^{+}_2=\left(L^{-}_2\right)^{\dagger}$ are defined by
\begin{align}
L^{-}_2&=\frac{d^2}{dx^2}+\eta\left(x\right)\frac{d}{dx}+\gamma\left(x\right), \label{2.10} \\
L^{+}_2&=\frac{d^2}{dx^2}-\eta\left(x\right)\frac{d}{dx}+\gamma\left(x\right)-\eta'\left(x\right),
\nonumber
\end{align}
with $\eta$ and $\mathscr{E}$ being given by 
\begin{equation}
\mathscr{E}=\frac{2m^{*}E}{\hbar^{2}},\qquad\eta\left(x\right)=2\left(k+\frac{e}{c\hbar}\mathcal{A}(x)\right).
\label{2.11}
\end{equation}
Up to here the function $\gamma$ is related with $\eta$, thus with $\mathcal{A}$, through 
$\gamma = \eta'/2 + \eta^2/4$, which is consistent with the fact that in the approximation leading to 
equation~(\ref{2.6}) the second-order intertwining operator $L^{-}_2$ is the square of a first-order 
operator, $L^{-}_2 = (\frac{d}{dx} + \frac{\eta}{2})^2$. In order to widen options, which includes the 
possibility that $L^{-}_2$ would be the product of two in general different first-order differential 
intertwining operators, from now on we will assume two things: first of all, $\eta$ and $\mathcal{A}$ 
will be always related by equation~(\ref{2.11}); on the other hand, the relation between $\gamma$ and 
$\eta$ is such that $L^{\pm}_2$ intertwine two Hermitian Schr\"odinger 
Hamiltonians \cite{6} (see equation~(\ref{2.17}) below). These mathematical assumptions have to do 
physically with the inclusion in the effective Hamiltonian of equation~(\ref{2.5}) of extra terms, 
that could be associated to large-distance hopping processes, spatially varying external potentials, 
etcetera \cite{1,12,CDO2020}.

Now, it is straightforward to decouple the system of equations (\ref{2.8}-\ref{2.9}) by applying $L_2^{+}$ and $L_2^{-}$ respectively, thus we get
\begin{align}
L_2^{+}L_2^{-}\psi^{(0)}(x)&=\mathscr{E}^{2}\psi^{(0)}(x), \nonumber\\
L_2^{-}L_2^{+}\psi^{(2)}(x)&=\mathscr{E}^{2}\psi^{(2)}(x). \label{2.12}
\end{align}
Note that $L_2^{+}L_2^{-}$  and $L_2^{-}L_2^{+}$ are fourth-order differential hermitian operators. More\-over, $\eta(x)$ is related linearly with the vector potential amplitude $\mathcal{A}(x)$, hence with the 
magnetic field as \cite{7,cf20}:
\begin{equation} \label{2.13}
\mathcal{B}(x)=\frac{c\hbar}{2e}\eta'(x).
\end{equation}
Another important point is that the second-order operators $L_2^{-}$ and $L_2^{+}$ transform $\psi^{(0)}(x)$ 
into $\psi^{(2)}(x)$ and vice versa. Thus, based on Eqs.~(\ref{2.6}-\ref{2.12}) it seems natural trying to 
adapt the second-order supersymmetric quantum mechanics \cite{ais93,aicd95,sa99,6,fe10,fe19} in the study of 
bilayer graphene in external magnetic fields.

\subsection{Second-order SUSY QM}

Let us assume that $\psi^{\left(0\right)}\left(x\right)$ and $\psi^{\left(2\right)}\left(x\right)$ are eigenfunctions of the non-relativistic Hamiltonians $H_{0}$ and $H_{2}$ respectively, which are given by
\begin{align}
H_{0}&=-\frac{d^{2}}{dx^2}+V_{0}(x),\nonumber\\
H_{2}&=-\frac{d^{2}}{dx^2}+V_{2}(x),
\label{2.14} 
\end{align}
where the so-called SUSY partner potentials $V_{0}$ and $V_{2}$ are to be determined. We will assume that an intertwining relation involving the two Hamiltonians and the operator in Eq.~(\ref{2.10}) is also fulfilled \cite{6}:
\begin{equation}
H_{2}L_2^{-}=L_2^{-}H_{0}.
\label{2.15}
\end{equation}
Taking this into account, after some work the following expressions are obtained:
\begin{align}
V_{2}\left(x\right)&=V_{0}\left(x\right)+2\eta'\left(x\right),\label{2.16}\\
\gamma\left(x\right)&=\frac{\eta^2\left(x\right)}{2}-\frac{\eta'\left(x\right)}{2}-V_{0}\left(x\right)+\frac{\epsilon_1+\epsilon_2}{2},\label{2.17}\\
V_{0}(x)&=\frac{\eta''\left(x\right)}{2\eta(x)}-\frac{\left(\eta'\left(x\right)\right)^2}{4\eta^{2}\left(x\right)}-\eta'\left(x\right)+\frac{\eta^2\left(x\right)}{4}+\bigg(\frac{\epsilon_1+\epsilon_2}{2}\bigg)+\bigg(\frac{\epsilon_1-\epsilon_2}{2\eta\left(x\right)}\bigg)^2\label{2.18},
\end{align}
with $\epsilon_1$ and $\epsilon_2$ being in general arbitrary complex numbers called {\it factorization energies}. In this paper we are going to choose them real, asking as well that $V_{0}$ and $V_{2}$ will be {\it shape invariant} SUSY partner potentials with known analytic solutions \cite{cr00,ba01,do07,8,gmr18,ju19}. 

Let us suppose now that the normalized eigenfunctions $\psi_n^{\left(0\right)}\left(x\right)$ and eigenvalues $\mathcal{E}_n^{(0)}$ of $H_0$ are given. Equation~(\ref{2.15}) implies that the eigenfunctions of $H_2$ can be found by acting $L_2^-$ onto $\psi_n^{\left(0\right)}$. In order to determine $L_2^-$, two {\it seed solutions} $u_1, \ u_2$ in the kernel of $L_2^-$ ($L_2^-u_1= L_2^-u_2 =0$) satisfying as well the stationary Schr\"odinger equation ($H_0 u_i = \epsilon_i u_i, \ i=1,2$) are required. Depending from the choice of $u_1, \ u_2$, $\epsilon_1, \ \epsilon_2$, we can get different variants for the spectrum of $H_2$ \cite{6}. In this paper we will restrict ourselves to the simplest possibility leading to a non-singular $V_2$, which consists in taking $u_1, \ u_2$ as the eigenstates of $H_0$ associated to the two lowest eigenvalues, $\epsilon_1=\mathcal{E}_1^{(0)}, \ \epsilon_2=\mathcal{E}_0^{(0)}$. With this choice, the spectrum of $H_0$ will have two extra eigenvalues $\mathcal{E}_0^{(0)}$ and $\mathcal{E}_1^{(0)}$ as compared with the spectrum of $H_2$, since both levels are deleted from Sp$(H_0)$ in order to create $H_2$. We summarize this by expressing now the normalized eigenstates of $H_2$, and associated eigenvalues, in terms of the corresponding ones of $H_0$:
\begin{eqnarray} \label{2.19}
\psi^{\left(2\right)}_{n}\left(x\right) = \frac{L_2^-\psi^{\left(0\right)}_{n+2}\left(x\right)}{\sqrt{(\mathcal{E}_{n+2}^{(0)}-\mathcal{E}_0^{(0)}) (\mathcal{E}_{n+2}^{(0)}-\mathcal{E}_1^{(0)})}}, \qquad \mathcal{E}_n^{(2)} = \mathcal{E}_{n+2}^{(0)}, \qquad n=0,1,\dots
\end{eqnarray}

Going back to our main task, the determination of the eigenstates and eigenvalues of the effective Hamiltonian $H$ for the electron in bilayer graphene, let us note first of all that
\begin{align}
L_2^{-}L_2^{+}\psi_n^{(2)}(x)&=(H_2-\epsilon_1)(H_2-\epsilon_2)\psi_n^{(2)}(x)
=\left(\mathcal{E}_n^{(2)}-\mathcal{E}_0^{(0)}\right)\left(\mathcal{E}_n^{(2)}-\mathcal{E}_1^{(0)}\right)\psi_n^{(2)}(x),\nonumber\\
L_2^{+}L_2^{-}\psi_n^{(0)}(x)&=(H_0-\epsilon_1)(H_0-\epsilon_2)\psi_n^{(0)}(x)
=\left(\mathcal{E}_n^{(0)}-\mathcal{E}_0^{(0)}\right)\left(\mathcal{E}_n^{(0)}-\mathcal{E}_1^{(0)}\right)\psi_n^{(0)}(x).
\label{2.20}
\end{align}  
Thus, the normalized eigenfunctions and eigenvalues of $H$ in terms of the ones of the auxiliar
Hamiltonians $H_0, \ H_2$ are given by
\begin{align}
\Psi_{0,j}(x,y)&=e^{iky}
\begin{pmatrix} 
0\\ 
\psi^{\left(0\right)}_{j}(x)
\end{pmatrix}, \qquad E_{0}=0, \qquad j=0,1,
\label{2.21} \\
\Psi_{n-1}\left(x,y\right)&=\frac{e^{iky}}{\sqrt{2}}
\begin{pmatrix} 
\psi^{\left(2\right)}_{n-2}\left(x\right)\\ 
\psi^{\left(0\right)}_{n}\left(x\right)
\end{pmatrix},\quad E_{n-1}=\pm\frac{\hbar^{2}}{2m^*}\sqrt{(\mathcal{E}_{n}^{(0)}-\mathcal{E}_0^{(0)})(\mathcal{E}_{n}^{(0)}-\mathcal{E}_1^{(0)})}, \quad n=2,3,\dots,
\label{2.22} 
\end{align}
where the positive eigenvalues are associated to electrons and the negative ones to holes. Note that the ground state has an extra index to indicate its double degeneracy.

In the next section we are going to analyze different kinds of magnetic fields and physical quantities, as the probability density $\rho=\Psi^{\dagger}\Psi$. If the state of the system is one of the eigenfunctions of the Hamiltonian $H$ it is obtained 
\begin{align}
\rho_{0,j}(x)&=\Psi_{0,j}^{\dagger}\Psi_{0,j}=|\psi^{(0)}_{j}|^{2},\quad j=0,1, \label{2.23}\\
\rho_{n+1}(x)&=\Psi_{n+1}^{\dagger}\Psi_{n+1}=\frac{1}{2}\{|\psi^{(2)}_{n}|^{2}+|\psi^{(0)}_{n+2}|^{2}\},\quad n=0,1,\dots
\label{2.24}
\end{align}
In addition, the current density is calculated through the usual procedure leading to\cite{9}
\begin{eqnarray}
& J_{\ell,n}=\frac{\hbar}{m^{*}}{\rm Im}\left(\Psi^{\dagger}_n \, j_{\ell} \, \Psi_n\right), \quad \ell=x,y, \label{2.25} \\
& j_{x}=\sigma_x\partial_x+\sigma_y\partial_y,\quad
j_{y}=\sigma_y\partial_x-\sigma_x\partial_y.\label{2.26}
\end{eqnarray}
In particular, for the two orthogonal eigenstates associated to the ground state energy of $H$ it turns out that
\begin{equation}
J_{x,0}=J_{y,0}=0.\label{2.27}
\end{equation}
On the other hand, for the excited states of $H$ the $x$ component is given by
\begin{equation}
J_{x,n+1}(x)=\frac{\hbar}{2m^{*}}{\rm Im}\left[W({\psi_{n+2}^{(0)}}^{*},\psi_{n}^{(2)})
+2k\psi_{n+2}^{(0)}{\psi_{n}^{(2)}}^{*}\right],\quad n=0,1,\dots \label{2.28}
\end{equation}
while the $y$ component becomes
\begin{equation}
J_{y,n+1}(x)=\frac{\hbar}{2m^{*}}{\rm Re}\left[W({\psi_{n+2}^{(0)}}^{*},\psi_{n}^{(2)})-2k\psi_{n+2}^{(0)}{\psi_{n}^{(2)}}^{*}\right],\quad n=0,1,\dots\label{2.29}
\end{equation}
where $W(f,g)=fg'- f'g$. It is straightforward to check that, if both $\psi^{(0)}$ and $\psi^{(2)}$ are real, then 
\begin{equation} \label{2.30}
J_{x,n+1}(x)=0, \quad J_{y,n+1}(x)=\frac{\hbar}{2m^{*}}\left[W(\psi_{n+2}^{(0)},\psi_{n}^{(2)})-2k\psi_{n+2}^{(0)}\psi_{n}^{(2)}\right].
\end{equation}

It is worth to stress that although the eigenvectors $\Psi(x,y)$ of equations (\ref{2.21}) and (\ref{2.22}) have as vector entries the eigenfunctions of the auxiliary potentials $V_2(x)$ and $V_0(x)$, these potentials are just mathematical tools useful to find exact analytic solutions to the eigenvalue problem posed in (\ref{2.6}). Thus, they should not be confused with physical potentials used to describe the interactions among an electron and the other components of the graphene layer, for which a Hamiltonian more general than (\ref{2.5}) could be employed \cite{Schutt2011}. In addition, the relation between the external magnetic fields involved implicitly in (\ref{2.8}) and the auxiliary potentials given by equations (\ref{2.13}) and (\ref{2.16}) is as well purely mathematical, since $V_2(x)$ and $V_0(x)$ do not have a direct physical meaning. However, it should be recalled that the magnetic fields, we are dealing with, are applied externally to the graphene layers. Then, despite it could be difficult to produce them nowadays in the laboratory there exist previous experimental studies on inhomogeneous magnetic fields applied to systems similar to the bilayer graphene that could be the basis for designing the magnetic profiles that we will examine here. Such inhomogeneous magnetic fields have been realized through magnetic vortexes \cite{Masir2011}, have used magnetic field spectroscopy devices \cite{Schnez2009}, and even a single barrier has been performed by ferromagnetic stripes \cite{Masir2008, Matulis1994}. We must also note that the phenomena we are going to describe in this work are similar to the ones occurring in graphene when it is deformed by strain or those in which the graphene is immersed in crossed external electric and magnetic fields. Even though these two processes and graphene under external magnetic fields represent phenomena of different physical nature, they are mathematically similar \cite{Maurice2015,Naumis2017,CDO2020}.
\newpage
\section{Solvable cases}

We will analyze now some special magnetic fields leading to pairs of auxiliar shape invariant SUSY partner potentials, which will supply us exact solutions to our original problem. Hereafter, the parameters $\omega$, $\alpha$, and $D$ appearing in our expressions will be taken positive\cite{10}. 

\subsection{Case I: constant magnetic field}
The first case to be analyzed is a constant magnetic field $\vec{B}=(0,0,B_0)$, obtained from the vector potential
$\vec{A}=(0,xB_0,0)$ so that $\eta=2k+\omega x$, where $\omega=2eB_0/c\hbar$ is a constant with dimensions of (length)$^{-2}$. If the factorization energies are chosen as $\epsilon_2=0$ and $\epsilon_1=\omega$, the SUSY partner potentials become
\begin{align}
V_{0}(x)&=\frac{\omega^2}{4}\bigg(x+\frac{2k}{\omega}\bigg)^2-\frac{\omega}{2},\label{3.1.1}\\
V_{2}(x)&=\frac{\omega^2}{4}\bigg(x+\frac{2k}{\omega}\bigg)^2+\frac{3}{2}\omega.\label{3.1.2}
\end{align} 
We can see that $V_{0}(x)$ and $V_{2}(x)$ are just shifted harmonic oscillator potentials. The corresponding eigenvalues for the auxiliar Hamiltonians $H_{0}$ and $H_{2}$ are given by
\begin{equation}
\mathcal{E}_{0}^{(0)}=0, \quad \mathcal{E}_{1}^{(0)}=\omega, \quad \mathcal{E}_{n}^{(0)}=\mathcal{E}_{n-2}^{(2)}=n\omega, \quad n=2,3,\dots\label{3.1.3}
\end{equation}
The corresponding eigenfunctions are expressed in terms of Hermite polynomials as follows:
\begin{equation}
\psi_{n}^{(0)}(\zeta)=\psi_{n}^{(2)}(\zeta)=c_n e^{-\frac{1}{2}\zeta^2}H_{n}(\zeta),\label{3.1.4}  
\end{equation}
where $c_n$ is a normalization factor and $\zeta=\sqrt{\omega/2}(x+2k/\omega)$, $n=0,1,2,\dots$

In this case the eigenvalues $E_n$ for the electrons in bilayer graphene are given by
\begin{equation}
E_{n-1} =\frac{\hbar^{2}\omega}{2m^{*}}\sqrt{n(n-1)},\quad n=1,2,\dots 
\label{3.1.5}
\end{equation} 
It is important to stress that these eigenvalues do not depend of the wavenumber $k$, even though the eigenfunctions and the two auxiliar potentials do.

\begin{figure}[ht]
\centering
\includegraphics[width=8.2cm, height=5.5cm]{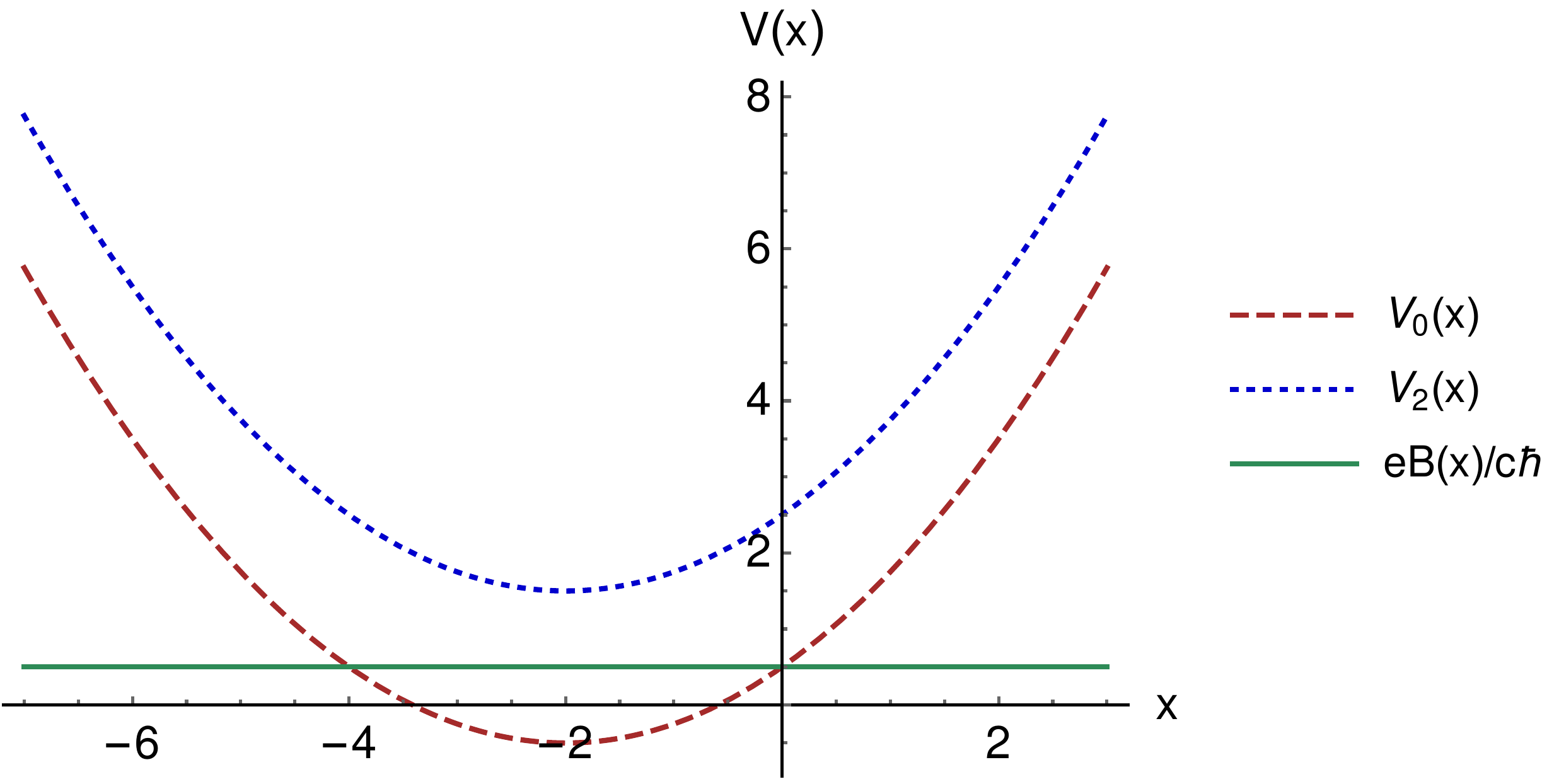}
\includegraphics[width=8.2cm,height=5.5cm]{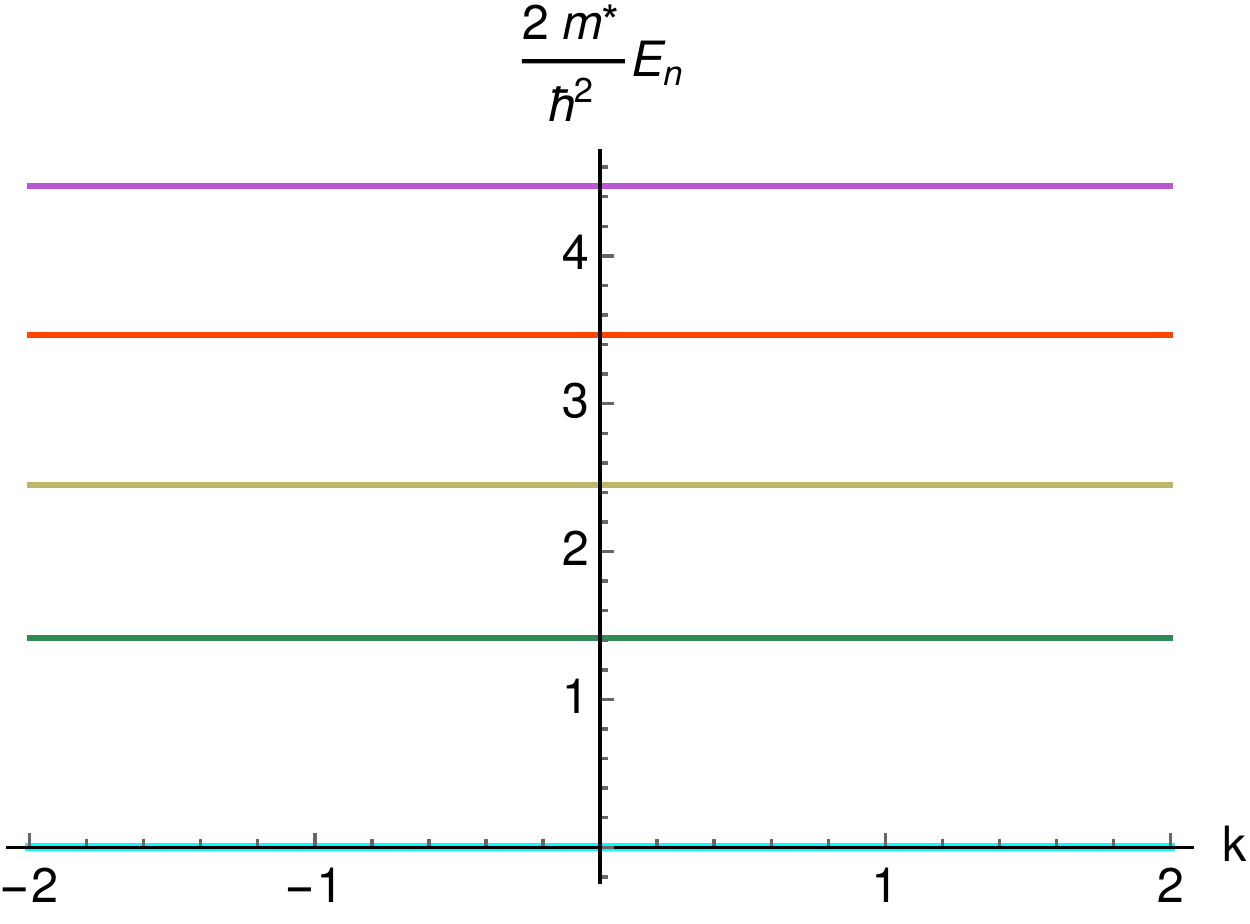}
\caption{(left) Plot of the potentials and the constant magnetic field as functions of $x$; (right) some eigenvalues $E_n$ as functions of $k$ for $\omega =1$ and $k=1$.}
\label{Imagen3_1_1}
\end{figure}

\begin{figure}[ht]
\centering
\includegraphics[width=8.2cm, height=5.5cm]{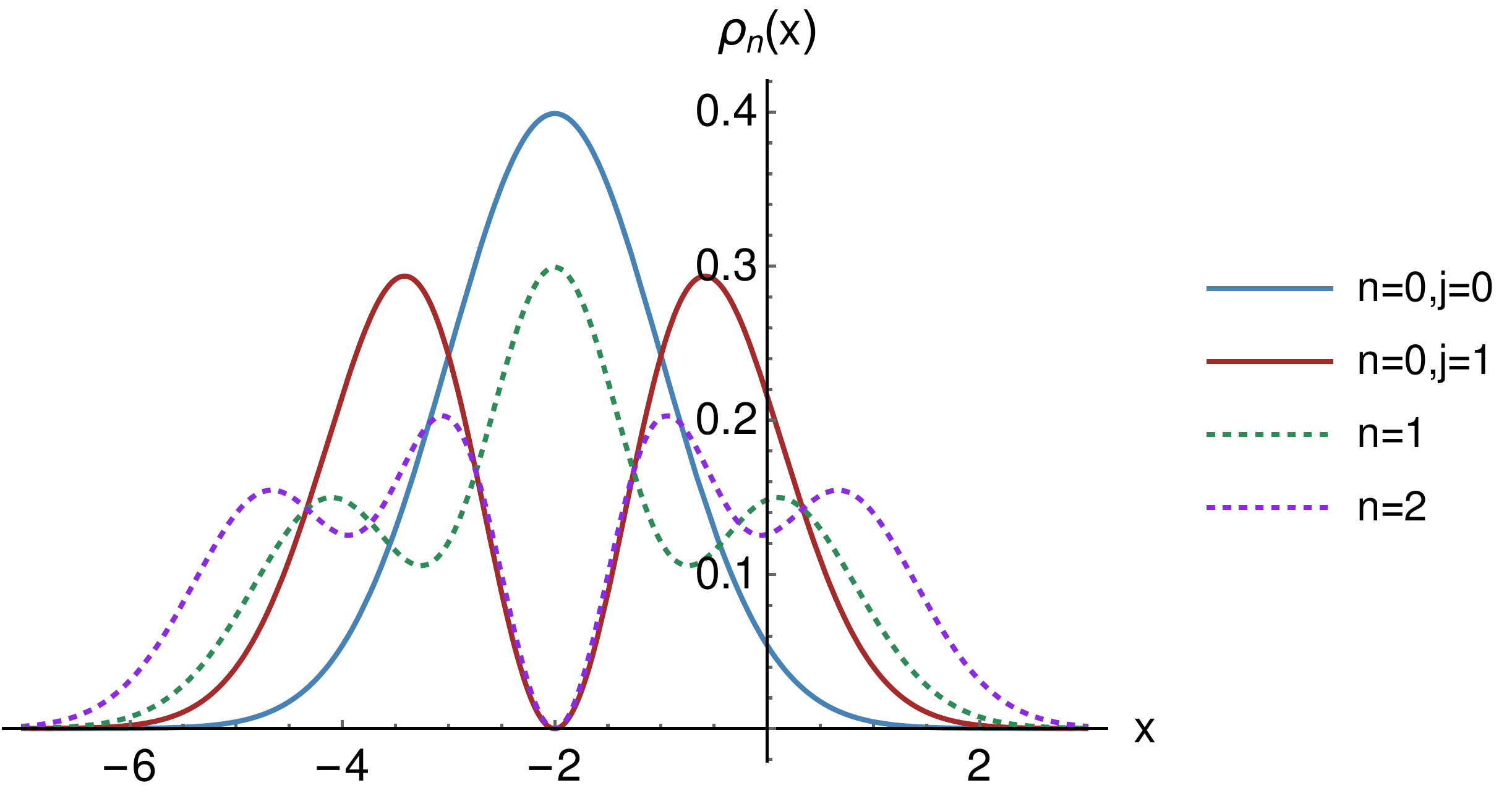}
\includegraphics[width=8.2cm, height=5.5cm]{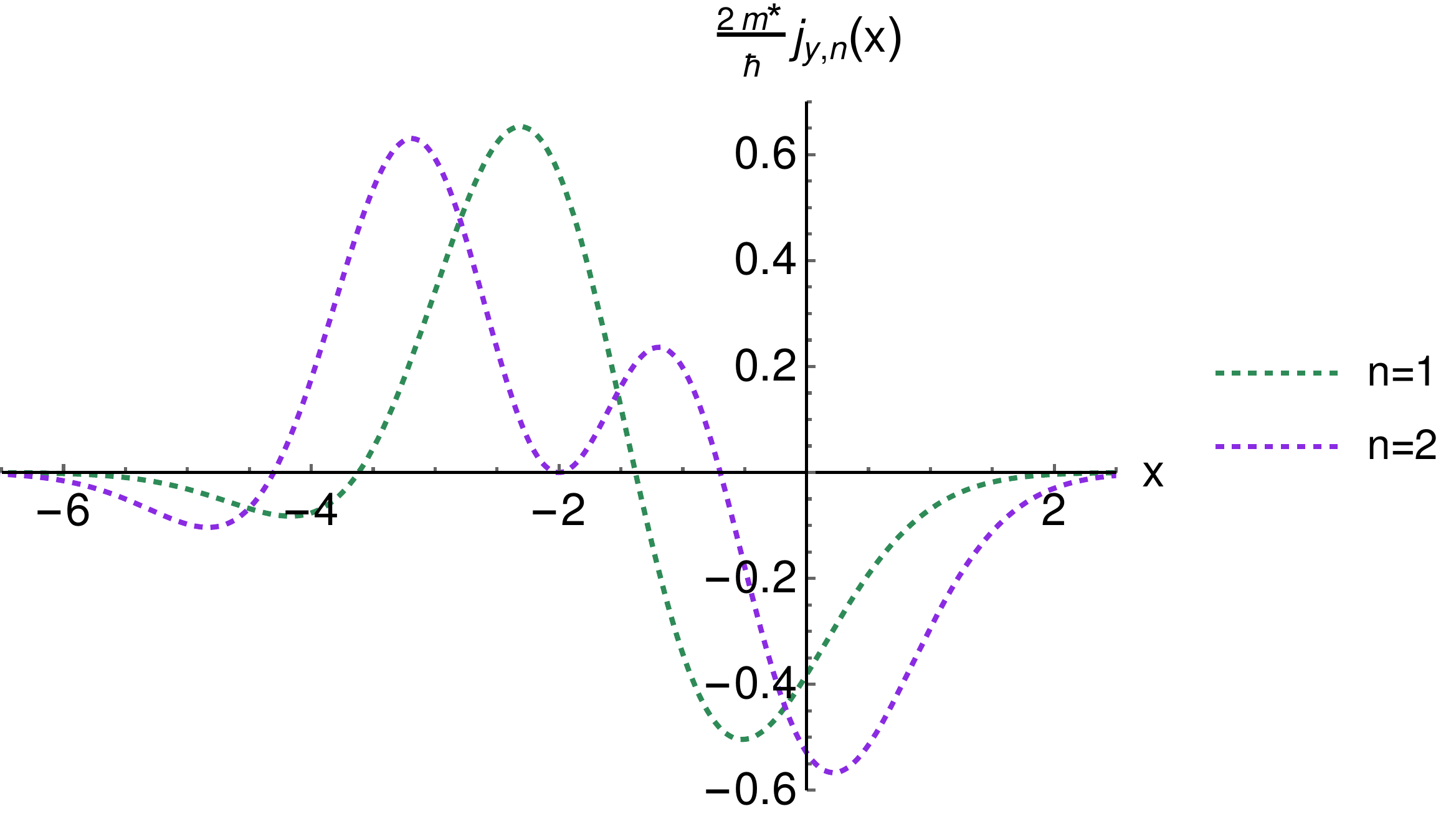}
\caption{(left) Plot of some probability densities for the constant magnetic field; (right) current densities for some eigenstates $\Psi_n(x)$ with $\omega=1$ and $k=1$.}  \label{Imagen3_1_2}
\end{figure}

Figure \ref{Imagen3_1_1} (a) shows plots of the potentials $V_0$, $V_2$ and the constant magnetic field leading to them, while Figure \ref{Imagen3_1_1} (b) sketches the first lowest eigenvalues $E_n$ as functions of $k$. Plots of the probability and current densities are shown in Figure \ref{Imagen3_1_2} (a and b, respectively).

\subsection{Case II: hyperbolic well}

The second case we are going to study is the magnetic field
\begin{equation}
\vec{B}=\left(0,0,B_0\textrm{sech}^{2}\alpha x\right)
\label{3.2.1}
\end{equation}
obtained from the vector potential $\vec{A}=\left(0,\frac{B_0}{\alpha}\textrm{tanh}\alpha x,0\right)$. According to equation~(\ref{2.11}), the $\eta$ function is now
\begin{equation}
\eta(x)=\left(2D-\alpha\right)\left(\frac{\kappa}{D-\alpha}+\textrm{tanh}\alpha x\right),
\label{3.2.2}
\end{equation}
where 
\begin{equation}
{D}=\frac{eB_{0}}{\hbar c\alpha}+\frac{\alpha}{2},\qquad \kappa=2k\left(\frac{{D}-\alpha}{2{D}-\alpha}\right)
\label{3.2.3}
\end{equation}
are constants having dimension of (length)$^{-1}$. In order to obtain auxiliar exactly solvable shape-invariant SUSY partner potentials, we need to choose the factorization energies as $\epsilon_2=0$ and 
$\epsilon_1={D}^{2}+\kappa^{2}-\left({D}-\alpha\right)^{2}-\frac{\kappa^{2}{D}^{2}}{\left({D}-\alpha\right)^{2}}$, so that
\begin{align}
V_{0}&={D}^{2}+\kappa^{2}-{D}({D}+\alpha)\textrm{sech}^{2}\alpha x+2\kappa{D}\textrm{tanh}\alpha x,\label{3.2.4}\\
V_{2}&={D}^{2}+\kappa^{2}-({D}-\alpha)({D}-2\alpha)\textrm{sech}^{2}\alpha x+2\kappa{D}\textrm{tanh}\alpha x, \label{3.2.5}
\end{align}
which are called Rosen-Morse II potentials in the literature. They will have a finite discrete spectrum for 
$|\kappa|<{D}$, with the eigenvalues of $H_0$ and $H_2$ being given by 
\begin{align}
& \mathcal{E}^{(0)}_{0}=0,\quad \mathcal{E}^{(0)}_{1}={D}^{2}+\kappa^{2}-\left({D}-\alpha\right)^{2}-\frac{\kappa^{2}{D}^{2}}{\left({D}-\alpha\right)^{2}},\nonumber\\
\mathcal{E}^{(0)}_{n}&=\mathcal{E}^{(2)}_{n-2}={D}^{2}+\kappa^{2}-\left({D}-n\alpha\right)^{2}-\frac{\kappa^{2}{D}^{2}}{\left({D}-n\alpha\right)^{2}}, \quad n=2,\dots,N,
\label{3.2.6}
\end{align}
where $N\alpha<{D}$. The corresponding eigenfunctions become
\begin{equation}
\psi_n^{(j)}(\zeta)=c_n\left(1-\zeta\right)^{\frac{s_j-n+a_j}{2}}\left(1+\zeta\right)^{\frac{s_j-n-a_j}{2}}\textrm{P}_n^{(s_j-n+a_j,s_j-n-a_j)}(\zeta),\quad
j=0,2, \ \ n=0,\dots,N.
\label{3.2.7}
\end{equation}
with $j=0,2,\ n=0,\dots,N.$ In this expression $c_n$ is a normalization factor, $\zeta=\textrm{tanh}\alpha x$, $s_0=\frac{{D}}{\alpha}$, $s_2=\frac{{D}}{\alpha}-2$,
$a_0=\frac{{D}\kappa}{\alpha({D}-n\alpha)}$, $a_2=\frac{{D}\kappa}{\alpha({D}-(n+2)\alpha)}$ and
$\textrm{P}_n^{(a,b)}(\zeta)$ are the Jacobi polynomials. In order to fulfill the square-integrability condition, the exponents of the first two 
factors in equation~(\ref{3.2.7}) need to be greater than zero. 

\begin{figure}[ht]
\centering
\includegraphics[width=8.2cm, height=5.5cm]{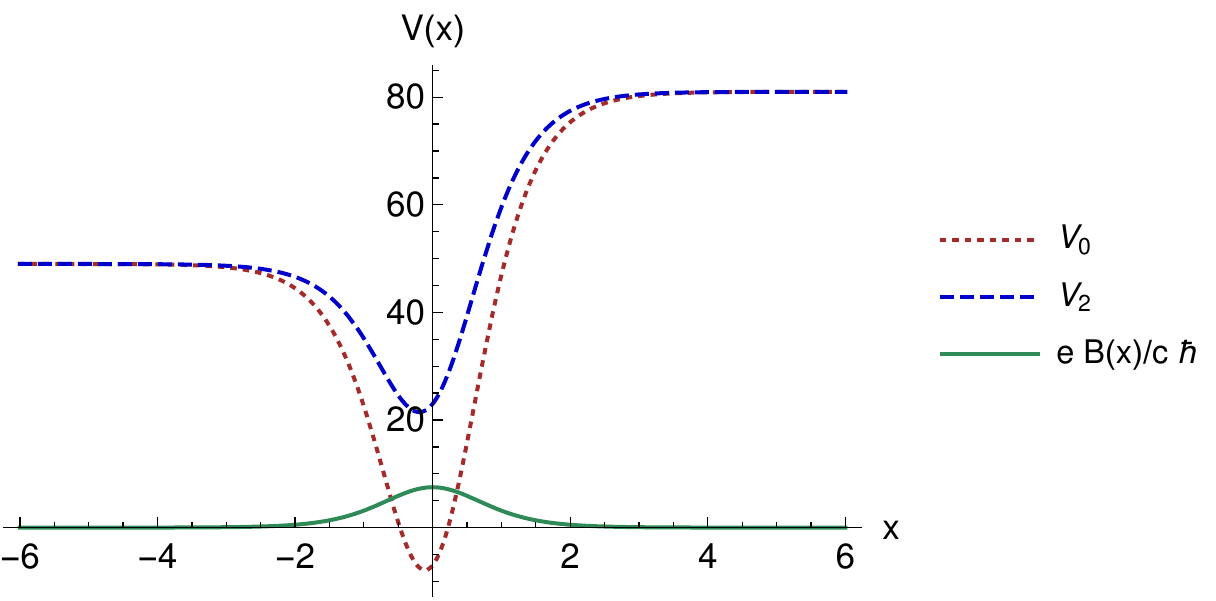}
\includegraphics[width=8.2cm, height=5.5cm]{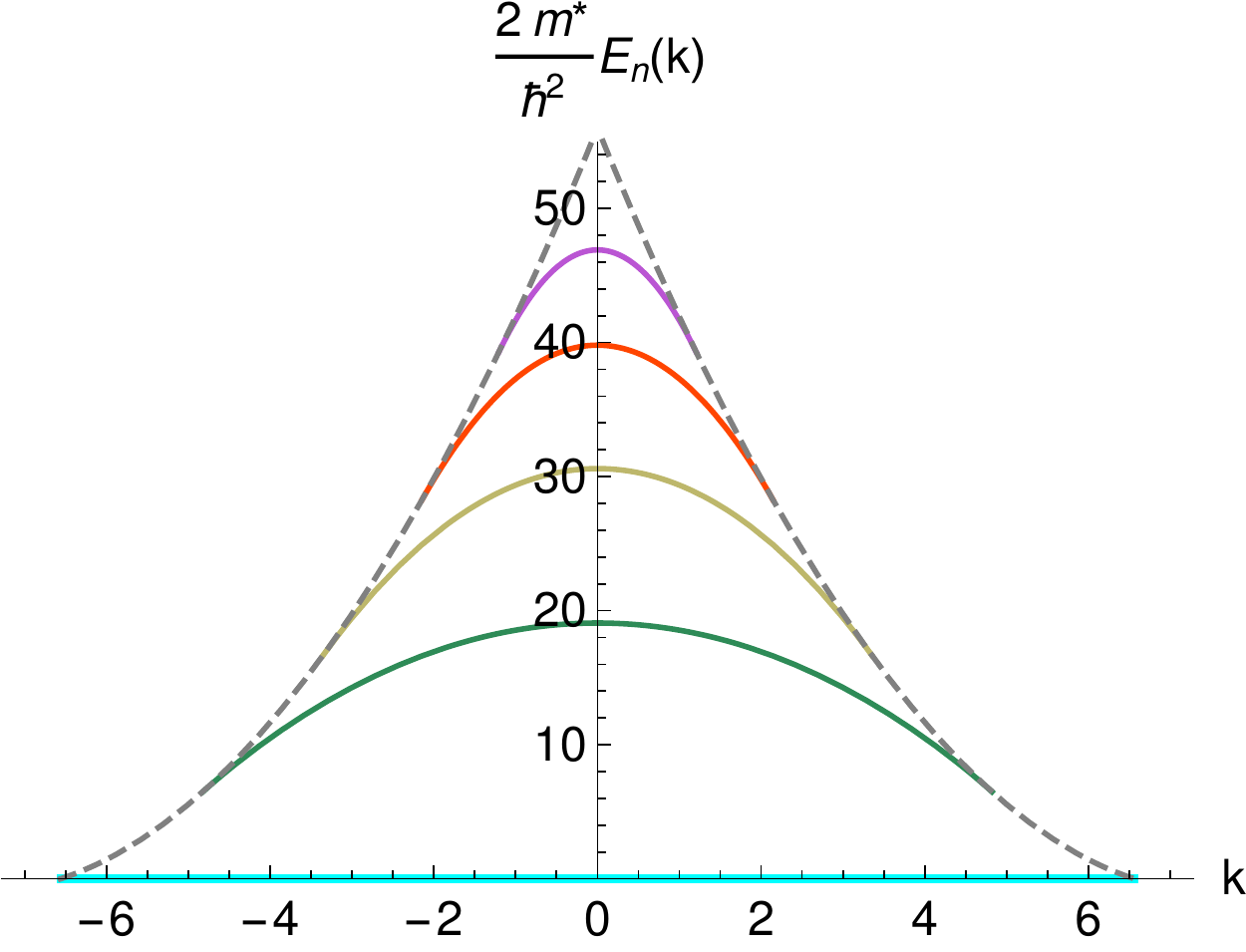}
\caption{Plot of the SUSY partner-potentials $V_0$, $V_2$ and the associated hyperbolic magnetic field (left); some eigenvalues $ E_n$ as functions of $k$ for ${D}=8$, $k=11/10$ and $\alpha=1$ (right).}
\label{Imagen3_2_1}
\end{figure}

The discrete eigenvalues of $H$ for electrons in this case are
\begin{equation}
E_{n-1}=\frac{\hbar^{2}}{2m^{*}}\mathcal{E}^{(0)}_{n}\sqrt{1-\gamma_n}, \ n=1,2,\dots,N,
\label{3.2.8}
\end{equation}
where
\begin{equation}
\gamma_n=\frac{{D}^{2}+\kappa^{2}-\left({D}-\alpha\right)^{2}-\frac{\kappa^{2}{D}^{2}}{\left({D}-\alpha\right)^{2}}}{{D}^{2}+\kappa^{2}-\left({D}-n\alpha\right)^{2}-\frac{\kappa^{2}{D}^{2}}{\left({D}-n\alpha\right)^{2}}}.
\label{3.2.9}
\end{equation}
Let us stress that these eigenvalues depend now on the wavenumber $k$. As can be seen in Figure \ref{Imagen3_2_1}, we have obtained a bounded finite discrete spectrum where an enveloping quadratic curve $ak^{2}+bk+c$ which touches the end points of $E_{n}$ can be drawn, with the constants $a$, $b$ and $c$ depending on the parameters $D$ and $\alpha$. The first derivative of this second degree polynomial is proportional to the group velocity in $y$-direction and the second derivative is a constant related to the component $\left[M_{inert}\right]_{2,2}$ of the effective mass tensor, i.e.,
\begin{equation}
\left[M_{inert}\right]_{2,2}=\frac{m^{*}}{a},\quad v_{g}=v_{F}^{2}\left(\frac{\hbar}{\gamma_1}\right)(2ak+b).
\label{3.2.10}
\end{equation}
For this specific case we have that $a=4D(D-\alpha)/(2D-\alpha)^{2}$, $b=2\alpha-4D^{2}/(2D-\alpha)$ and $c=D(D-\alpha)$. From now on we will write explicitly the constants $a$, $b$ and $c$ in the cases where it is possible.

\begin{figure}[ht]
\centering
\includegraphics[width=8.2cm, height=5.5cm]{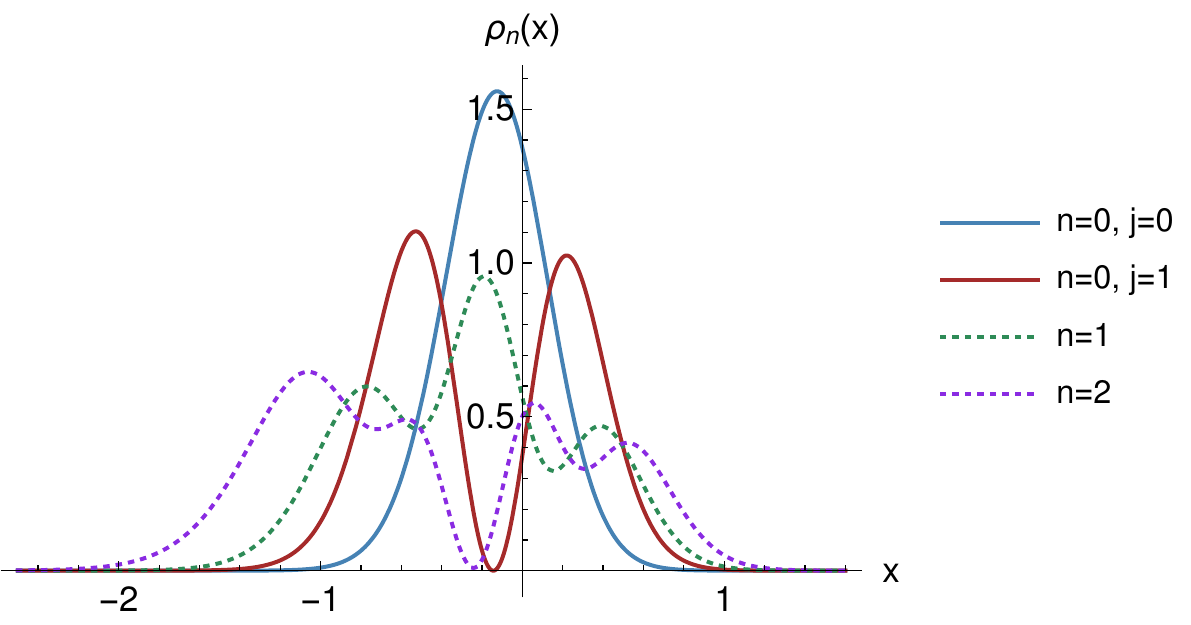}
\includegraphics[width=8.2cm, height=5.5cm]{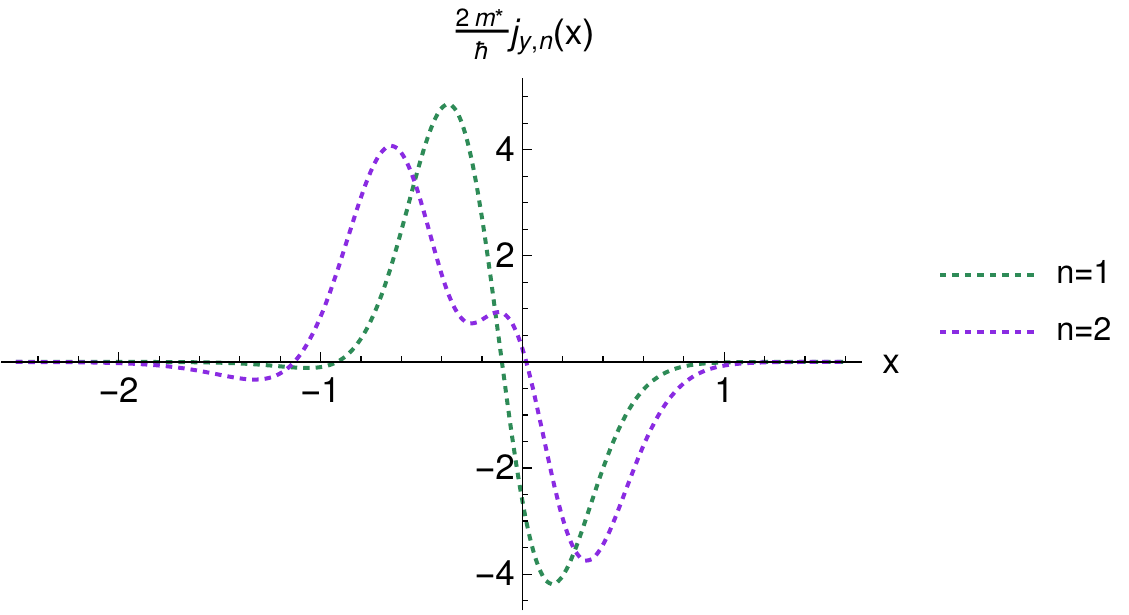}
\caption{Plots of some probability densities (left) and currents (right) for the hyperbolic well with ${D}=8$, $k=11/10$ and $\alpha=1$.}
\label{Imagen3_2_2}
\end{figure}

In Figure \ref{Imagen3_2_1} (a) we have drawn the potentials $V_0$, $V_2$ and the corresponding magnetic field, while the eigenvalues $E_n$ as functions of $k$ are shown in Figure \ref{Imagen3_2_1} (b). In Figure \ref{Imagen3_2_2} (a) the probability densities are sketched while Figure~\ref{Imagen3_2_2} (b) illustrates the probability currents for ${D}=8,\;\kappa=1,\;\alpha=1$.  

\subsection{Case III: trigonometric singular well}
In this case the magnetic field is given by
\begin{equation}
\vec{B}=(0,0,B_0\csc^2\alpha x), \quad 0\leq\alpha x\leq\pi,
\label{3.3.1}
\end{equation}
which is obtained from the vector potential $\vec{A}(x)=(0,-\frac{B_0}{\alpha}\cot\alpha x,0),$ and leads to
\begin{equation}
\eta=(2{D}+\alpha)\left(\frac{\kappa}{{D}+\alpha}-\cot\alpha x\right),
\label{3.3.2}
\end{equation}
where
\begin{equation}
{D}=\frac{eB_0}{c\hbar\alpha}-\frac{\alpha}{2},\qquad
\kappa=2k\left(\frac{{D}+\alpha}{2{D}+\alpha}\right),
\label{3.3.3}
\end{equation}
are two constants having dimension of (length)$^{-1}$. In order to get auxiliar shape-invariant SUSY partner potentials, we are going to take $\epsilon_2=0$ and $\epsilon_1=\kappa^2-{D}^2+({D}+\alpha)^2-\kappa^2{D^2}/({D}+\alpha)^2$. Thus
\begin{align}
V_{0}(x)&=\kappa^2-{D}^2+{D}({D}-\alpha)\csc^2\alpha x-2\kappa{D}\cot\alpha x,\label{3.3.4}\\
V_{2}(x)&=\kappa^2-{D}^2+({D}+2\alpha)({D}+\alpha)\csc^2\alpha x-2\kappa{D}\cot\alpha x,\label{3.3.5}
\end{align} 
which are the trigonometric Rosen-Morse (TRM) potentials. Note that it is possible to find in the literature the TRM potential expressed as $V_{-}(x)=A(A-1)\csc^2 x+2B\cot x-A^2+B^2/A^2$ (see \cite{gmr18}, page 56), which in principle admits bound state exact analytic solutions for $B\geq 0$, but also for $B < 0$ (see \cite{ju19}, page~4-12). Taking this into account, if in the expression for $V_{-}(x)$ we just choose $B=-kA$, we will recover precisely the TRM potentials of equations~(\ref{3.3.4}~, \ref{3.3.5}), which will be used here in order to keep consistency with \cite{10}.

The associated energy eigenvalues are
\begin{align}
& \mathcal{E}_{0}^{(0)}=0, \hskip1cm \mathcal{E}_{1}^{(0)}=\kappa^2-{D}^2+({D}+\alpha)^2-\frac{\kappa^2{D^2}}{({D}+\alpha)^2},\label{3.3.6}\\
\mathcal{E}_{n}^{(0)}&=\mathcal{E}_{n-2}^{(2)}=\kappa^2-{D}^2+({D}+n\alpha)^2-\frac{\kappa^2{D^2}}{({D}+n\alpha)^2}, \quad n=2,3,\dots,
\label{3.3.7}
\end{align}
while the corresponding eigenfunctions are expressed in terms of pseudo Jacobi polynomials as follows:
\begin{equation}
\psi_{n}^{(j)}=c_n (-1)^{-\frac{S_{j}+n}{2}}(\zeta^2+1)^{-\frac{S_{j}+n}{2}}e^{a_{j}\textrm{arccot}(\zeta)}P_{n}^{(-s_{j}-n-ia_{j},-s_{j}-n+ia_{j})}(i\zeta),  \label{3.3.8}
\end{equation}
where $j=0,2,\ n=0,1,2,\dots$, $c_n$ is a normalization factor, $s_{0}={D}/\alpha$, $s_{2}=s_{0}+2$, $a_{0}=\frac{-\kappa{D}}{\alpha ({D}+n\alpha)}$,
$a_{2}=\frac{-\kappa{D}}{\alpha ({D}+2\alpha+n\alpha)}$ and $\zeta=\cot\alpha x$. Notice that an alternative expression for the eigenfuntions of the trigonometric Rosen-Morse potential in terms of real orthogonal polynomials of real argument is given in \cite{11}.  

\begin{figure}[ht]
\centering
\includegraphics[width=8.2cm, height=5.5cm]{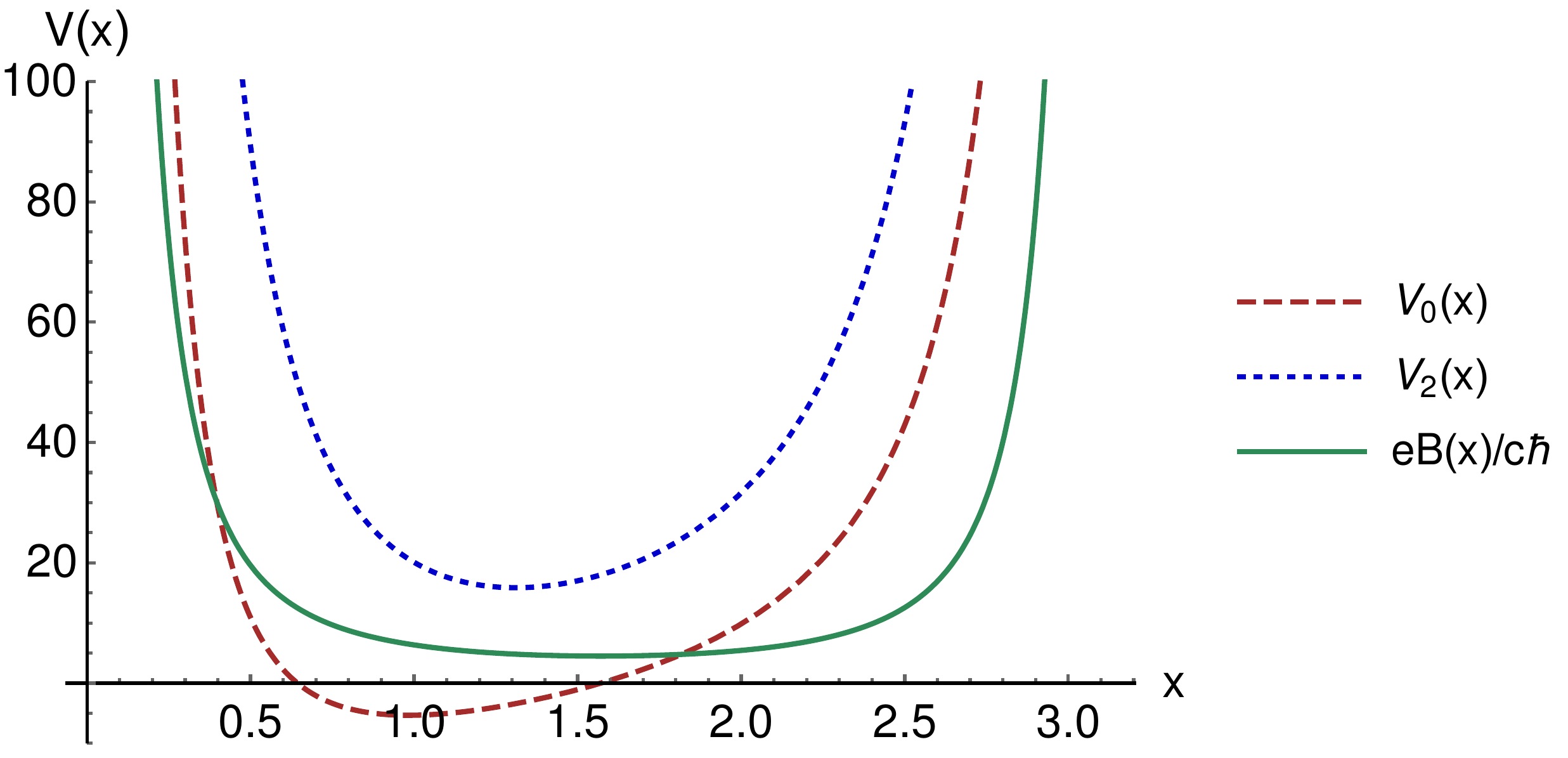}
\includegraphics[width=8.2cm, height=5.5cm]{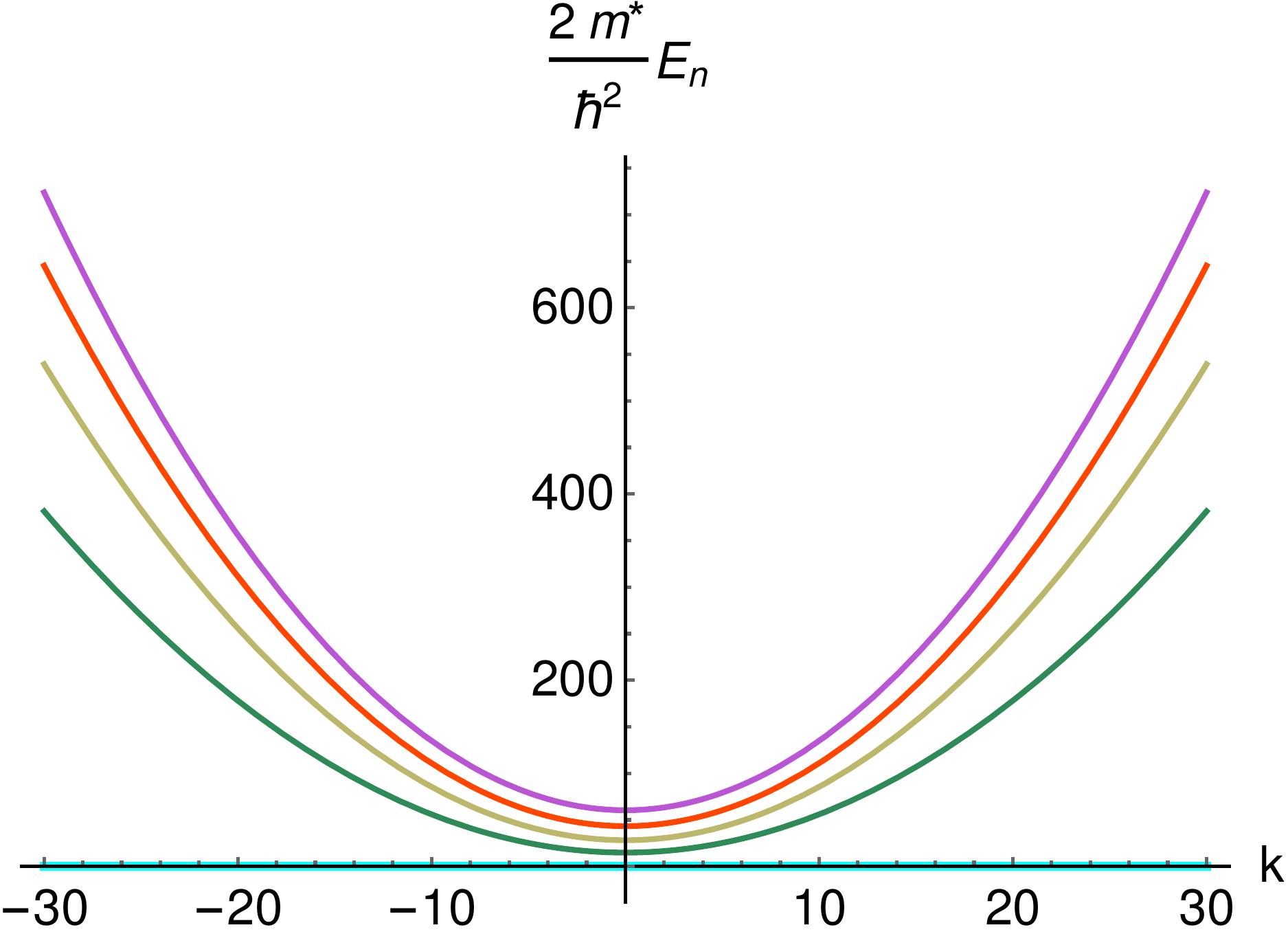}
\caption{(left) Plot of the potentials and magnetic field for the trigonometric singular well; (right) some eigenvalues $E_n$ as functions of $k$ for ${D}=4$, $k=9/5$ and $\alpha=1$.}
\label{Imagen3_3_1}
\end{figure}

\begin{figure}[ht]
\centering
\includegraphics[width=8.2cm, height=5.5cm]{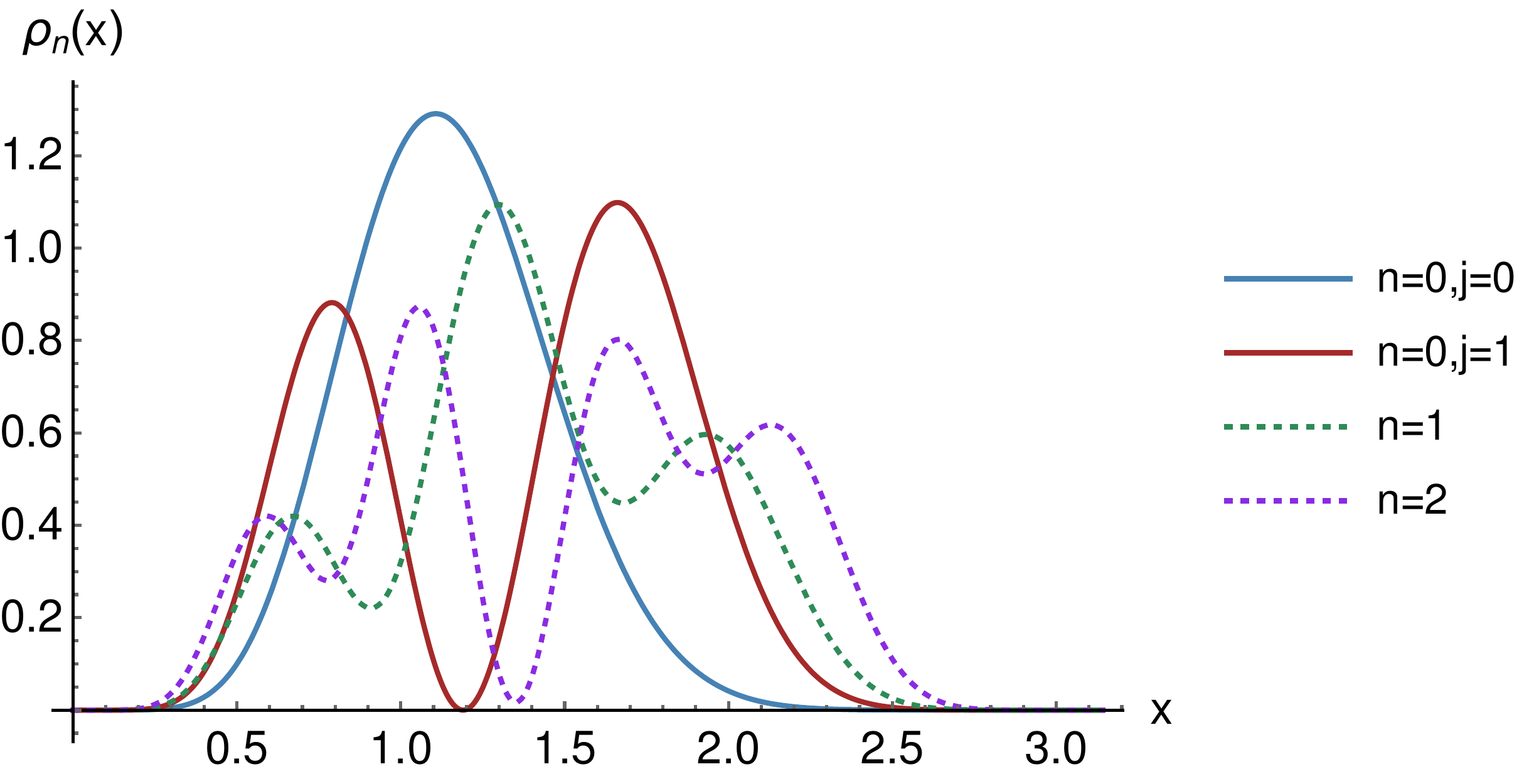}
\includegraphics[width=8.2cm, height=5.5cm]{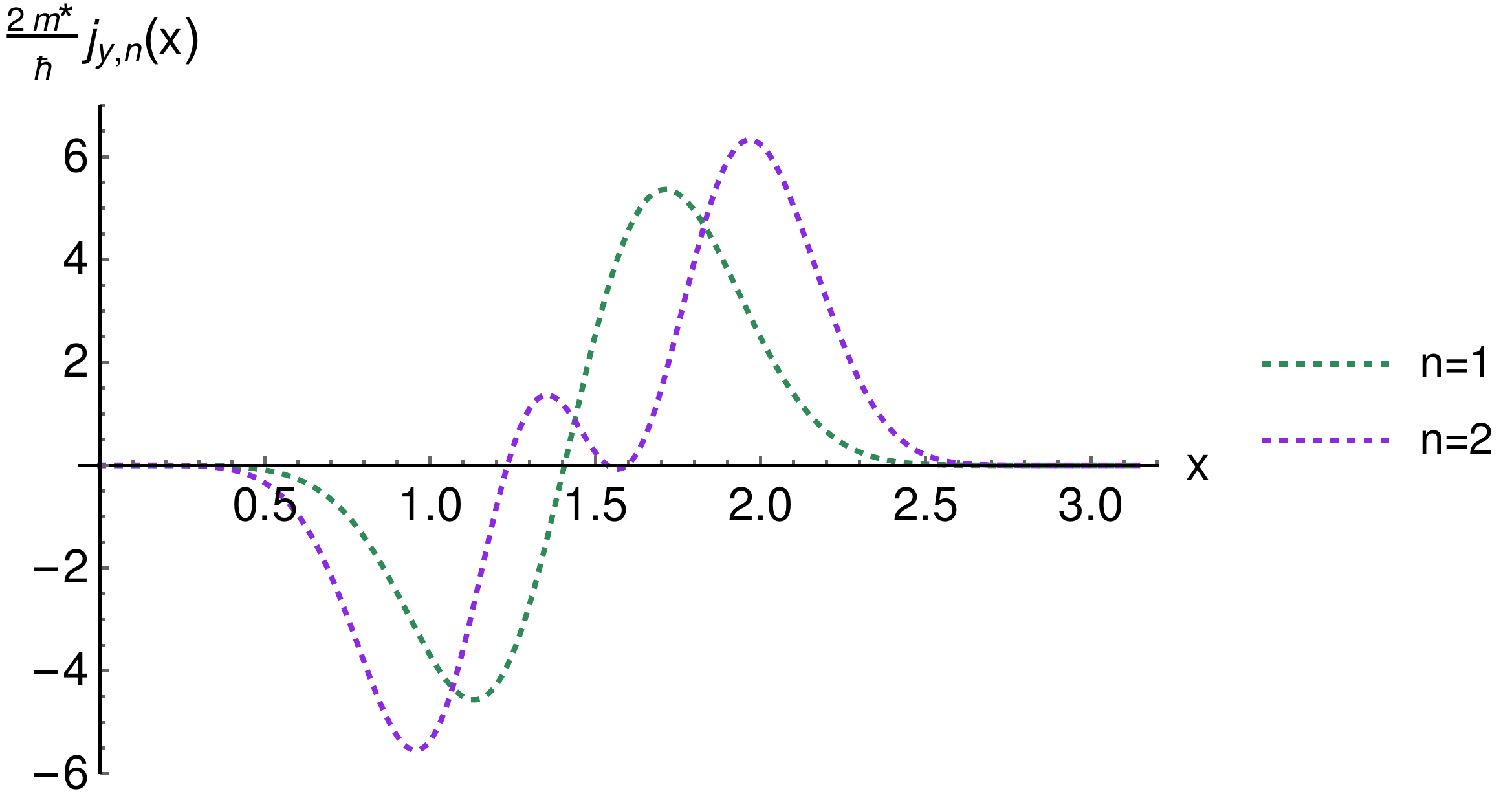}
\caption{Plots of some probability densities (left) and currents (right) for the trigonometric singular well with ${D}=4$, $k=9/5$ and $\alpha=1$.}
\label{Imagen3_3_2}
\end{figure}

The eigenvalues of the bilayer effective Hamiltonian $H$ for electrons are now
\begin{equation}
E_{n-1}=\frac{\hbar^{2}}{2m^{*}}\mathcal{E}^{(0)}_{n}\sqrt{1-\gamma_n}, \quad n=1,2,3,\dots
\label{3.3.9}
\end{equation}
where
\begin{equation}
\gamma_n=\frac{\kappa^2-{D}^2+({D}+\alpha)^2-\kappa^2{D^2}/({D}+\alpha)^2}{\kappa^2-{D}^2+({D}+n\alpha)^2-\kappa^2{D}^2/({D}+n\alpha)^2}.
\label{3.3.10}
\end{equation}
Once again, these eigenvalues depend on $k$, but now this dependence does not impose any restriction on them. Figure \ref{Imagen3_3_1} (a) shows a plot of the magnetic field and the potentials $V_0$, $V_2$ of equations (\ref{3.3.1}) and (\ref{3.3.4},\ref{3.3.5}) respectively, while Figure \ref{Imagen3_3_1} (b) sketches the first eigenvalues $E_n$ as functions of $k$. In Figure \ref{Imagen3_3_2} we have drawn the probability densities (a) and currents (b) for some eigenfunctions, associated to the lowest eigenvalues.

\subsection{Case IV: exponentially decaying magnetic field}
For a magnetic field $\vec{B}$ decaying in the positive $x$-direction in the way
\begin{equation}
\vec{B}=\left(0,0,B_0e^{-\alpha x}\right),
\label{4.4.1}
\end{equation}
the vector potential reads
\begin{equation}
\vec{A}=\left(0,-\frac{B_0}{\alpha}e^{-\alpha x},0\right).
\label{4.4.2}
\end{equation}
Thus, the key function $\eta$ becomes 
\begin{equation}
\eta(x)=2\kappa-\alpha-2{D}\,\exp\left(-\alpha x\right),
\label{4.4.3}
\end{equation}
where ${D}$ and $\kappa$ are constants with dimension of (length)$^{-1}$ given by
\begin{equation}
{D}=\frac{B_{0}e}{\hbar c\alpha},\qquad \kappa=k+\frac{\alpha}{2}.
\label{4.4.4}
\end{equation}
In order to get auxiliar exactly solvable shape-invariant SUSY partner potentials, we will choose the factorization energies as $\epsilon_2=0$ and $\epsilon_1=\kappa^{2}-(\kappa-\alpha)^{2}$, so that
\begin{align}
V_{0}&=\kappa^{2}+{D}^{2}e^{-2\alpha x}-2{D}\left(\kappa+\frac{\alpha}{2}\right)e^{-\alpha x} , \label{4.4.5}\\
V_{2}&=\kappa^{2}+{D}^{2}e^{-2\alpha x}-2{D}\left(\kappa-\frac{3\alpha}{2}\right)e^{-\alpha x} , \label{4.4.6}
\end{align}
which are the Morse potentials. Their corresponding eigenvalues are given by
\begin{eqnarray}
& \mathcal{E}^{(0)}_{0}=0,\quad \mathcal{E}^{(0)}_{1}=\kappa^{2}-(\kappa-\alpha)^{2},\nonumber\\
\mathcal{E}^{(0)}_{n}&=\mathcal{E}^{(2)}_{n-2}=\kappa^{2}-(\kappa-n\alpha)^{2}, \quad n=2,3,\dots
\label{4.4.7}
\end{eqnarray}
while the associated eigenfunctions are
\begin{equation}
\psi_n^{(j)}(\zeta)=c_n \zeta^{s_j-n}e^{-\frac{\zeta}{2}}\textrm{L}_n^{2s_j-2n}(\zeta),\quad j=0,2, \ n=0,1,\dots
\label{4.4.8}
\end{equation}
In equation~(\ref{4.4.8}) $c_n$ is a normalization constant and $\textrm{L}_n^{a}$ are the associated Laguerre polynomials with  $s_0=\frac{\kappa}{\alpha}$, $s_2=\frac{\kappa}{\alpha}-2$ and $\zeta=\frac{2{D}}{\alpha}e^{-\alpha x}$. In order to fulfill the square-integrability condition it is necessary that $\kappa>n\alpha$. 

\begin{figure}[ht]
\centering
\includegraphics[width=8.2cm, height=5.5cm]{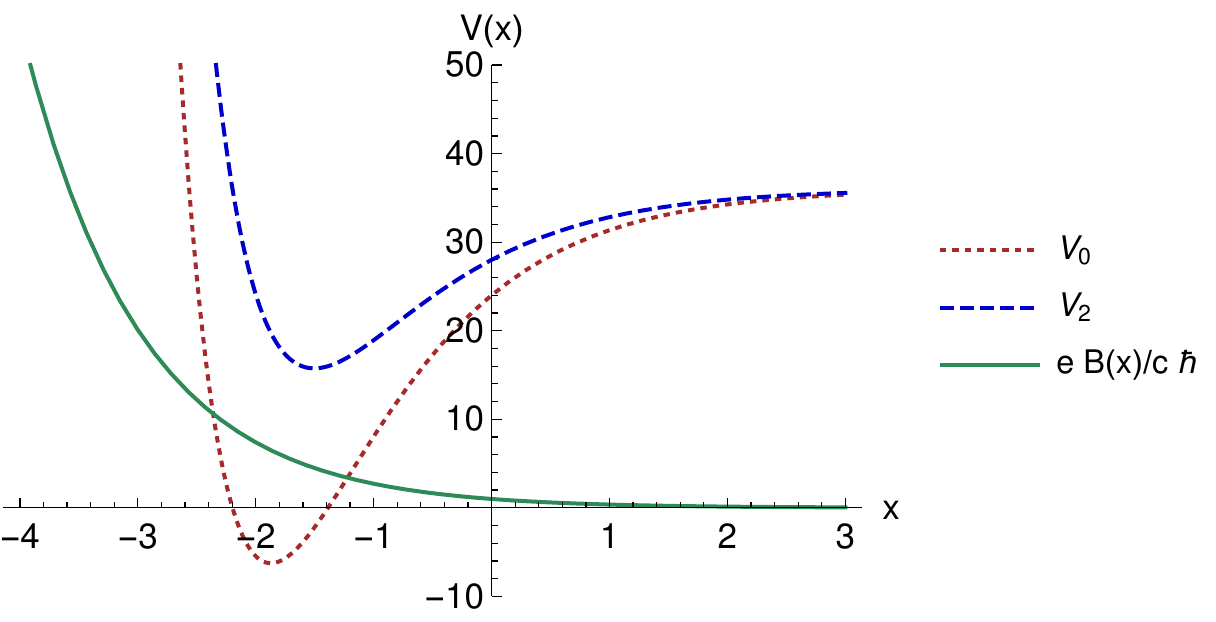}
\includegraphics[width=8.2cm, height=5.5cm]{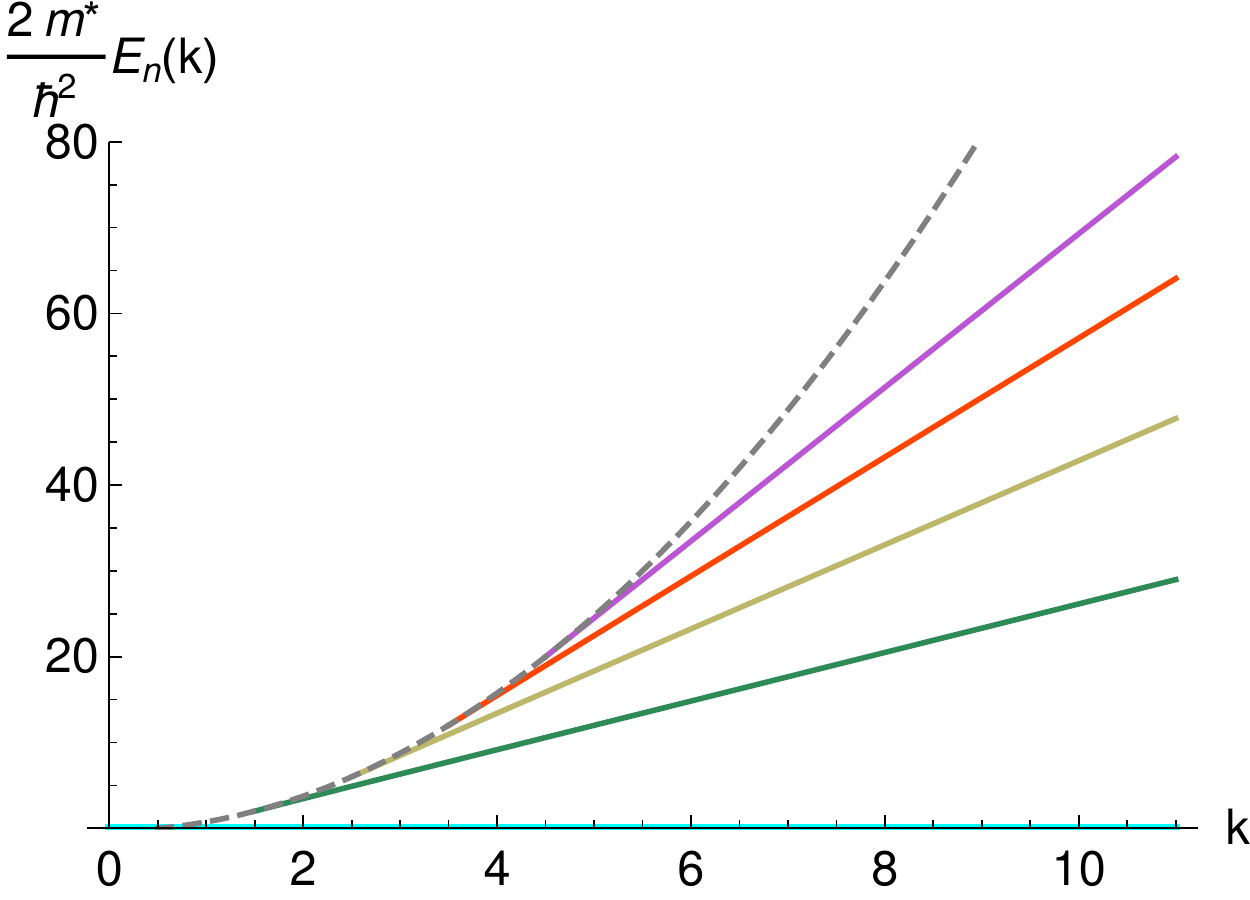}
\caption{Plot of the SUSY partner potentials $V_0$, $V_2$ and the exponentially decaying magnetic field (left) for ${D}=1$, $k=11/2$ and $\alpha=1$; some eigenvalues $ E_n$ as functions of $k$ (right).}
\label{Imagen3_4_1}
\end{figure}

\begin{figure}[ht]
\centering
\includegraphics[width=8.2cm, height=5.5cm]{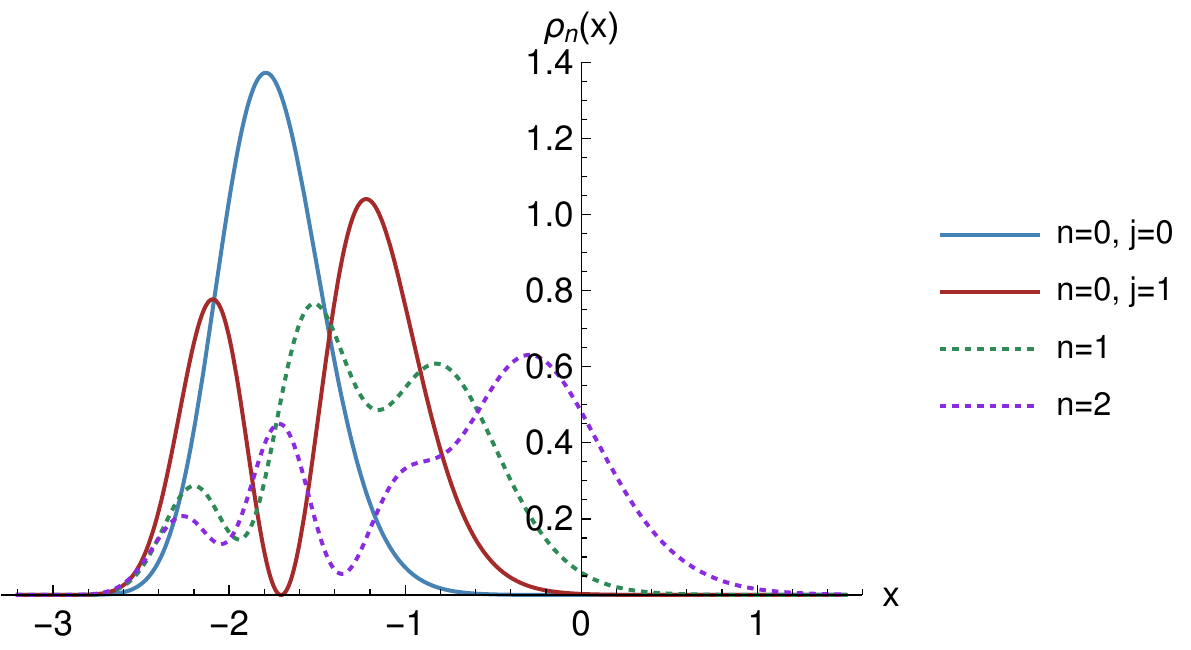}
\includegraphics[width=8.2cm, height=5.5cm]{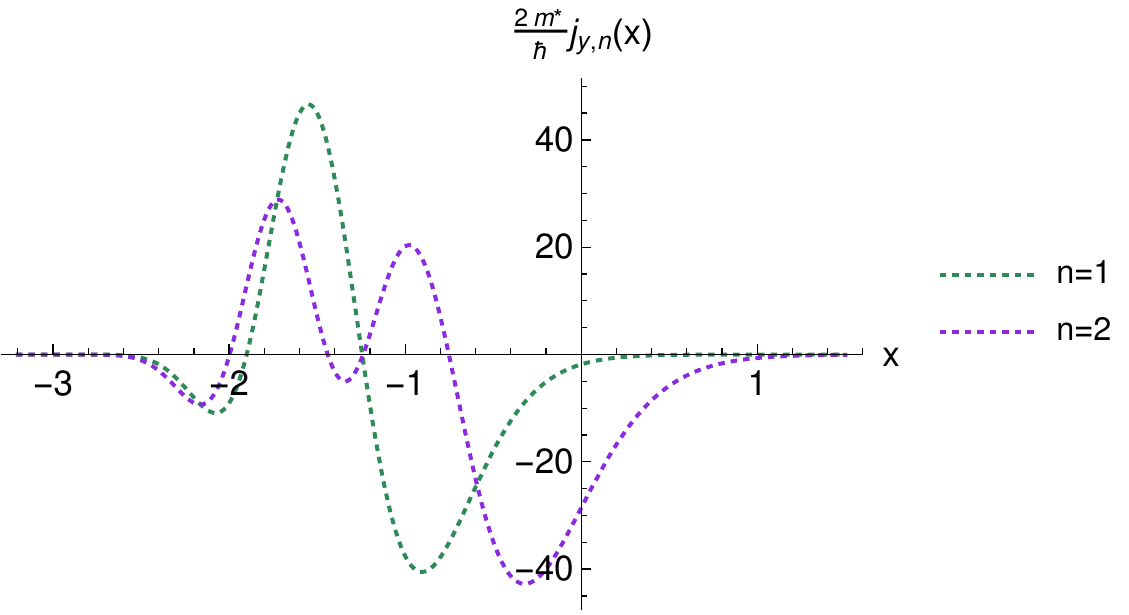}
\caption{Plots of some probability densities (left) and currents (right) for an exponentially decaying magnetic field with ${D}=1$, $k=11/2$ and $\alpha=1$.}  
\label{Imagen3_4_2}
\end{figure}

By collecting the previous information, the eigenvalues of $H$ for electrons in bilayer graphene become now
\begin{equation}
E_{n-1}=\frac{\hbar^{2}}{2m^{*}}\mathcal{E}^{(0)}_{n}\sqrt{1-\gamma_n},\quad \ n=1,2,3,\dots,
\label{4.4.9}
\end{equation} 
where
\begin{equation}
\gamma_n=\frac{\kappa^{2}-(\kappa-\alpha)^{2}}{\kappa^{2}-(\kappa-n\alpha)^{2}}.
\label{4.4.10}
\end{equation}
Notice that these energies do not depend of ${D}$, which is proportional to the magnetic field strength, although the eigenfunctions do. In addition, the parameters of the enveloping quadratic polynomial bounding the eigenenergies now are given by
\begin{equation}
a=1,\quad b=0,\quad c=-\frac{\alpha^{2}}{4}.
\label{4.4.11}
\end{equation}
Plots of the magnetic field, the potentials and the energy levels as functions of $k$ are shown in Figure \ref{Imagen3_4_1}. The corresponding probability and current densities are drawn in Figure \ref{Imagen3_4_2}.

\subsection{Case V: hyperbolic singular field}
Let us take now the magnetic field as 
\begin{equation}
\vec{B}=(0,0,B_0 \,\textrm{csch}^2\alpha x).
\label{3.5.1}
\end{equation}
The corresponding vector potential is given by $A(x)=(0,-(B_0/\alpha)\,\textrm{coth}\alpha x,0)$. Hence:
\begin{equation}
\eta=(2{D}+\alpha)\left(\frac{\kappa}{{D}+\alpha}-\textrm{coth}\alpha x\right),
\label{3.5.2}
\end{equation}
with 
\begin{equation}
{D}=\frac{eB_0}{c\hbar\alpha}-\frac{\alpha}{2},\qquad
\kappa=2k\left(\frac{{D}+\alpha}{2{D}+\alpha}\right),
\label{3.5.3}
\end{equation}
whose dimensions are (length)$^{-1}$. By choosing now $\epsilon_2=0$ and $\epsilon_1=\kappa^2+{D}^2-({D}+\alpha)^2-\frac{\kappa^2{D}^2}{({D}+\alpha)^2}$ we obtain the following auxiliar shape invariant SUSY partner potentials
\begin{align}
V_{0}(x)&=\kappa^2+{D}^2+{D}({D}-\alpha)\,\textrm{csch}^2\alpha x-2\kappa{D}\coth\alpha x,\label{3.5.4}\\
V_{2}(x)&=\kappa^2+{D}^2+({D}+2\alpha)({D}+\alpha)\,\textrm{csch}^2\alpha x-2\kappa{D}\coth\alpha x,\label{3.5.5}
\end{align}
which are called Eckart potentials in the literature. The eigenenergies become
\begin{align}
& \mathcal{E}_{0}^{(0)}=0,\quad
\mathcal{E}_{1}^{(0)}=\kappa^2+{D}^2-({D}+\alpha)^2-\frac{\kappa^2{D}^2}{({D}+\alpha)^2},\nonumber\\
\mathcal{E}_{n}^{(0)}&=\mathcal{E}_{n-2}^{(2)}=\kappa^2+{D}^2-({D}+n\alpha)^2-\frac{\kappa^2{D}^2}{({D}+n\alpha)^2}, \quad n=2,3,\dots
\label{3.5.6}
\end{align}
The corresponding eigenfunctions are expressed in terms of Jacobi polynomials as follows
\begin{equation}
\psi_{n}^{j}(\zeta)=c_n (\zeta-1)^{-\frac{s_{j}+n-a_{j}}{2}}(\zeta+1)^{-\frac{s_{j}+n+a_{j}}{2}}P_{n}^{(-s_{j}-n+a_{j},-s_{j}-n-a_{j})}(\zeta), \label{3.5.7}
\end{equation}
where $j=0,2, \ n=0,1,2,\dots$, $c_n$ is a normalization factor, $s_{0}=D/\alpha$, $s_{2}=s_{0}+2$, $a_{0}=\frac{\kappa{D}}{\alpha ({D}+n\alpha)}$,
$a_{2}=\frac{\kappa{D}}{\alpha ({D}+2\alpha+n\alpha)}$ and $\zeta=\coth\alpha x$. The boundary conditions for the eigenfunctions $\psi_{n}^{j}(\zeta)$ impose the constrain $\kappa>D>0$. In addition, the exponent of the first factor in equation~(\ref{3.5.7}) must be greater than zero and the second must be negative to satisfy the square-integrability condition.

\begin{figure}[ht]
\centering
\includegraphics[width=8.2cm, height=5.5cm]{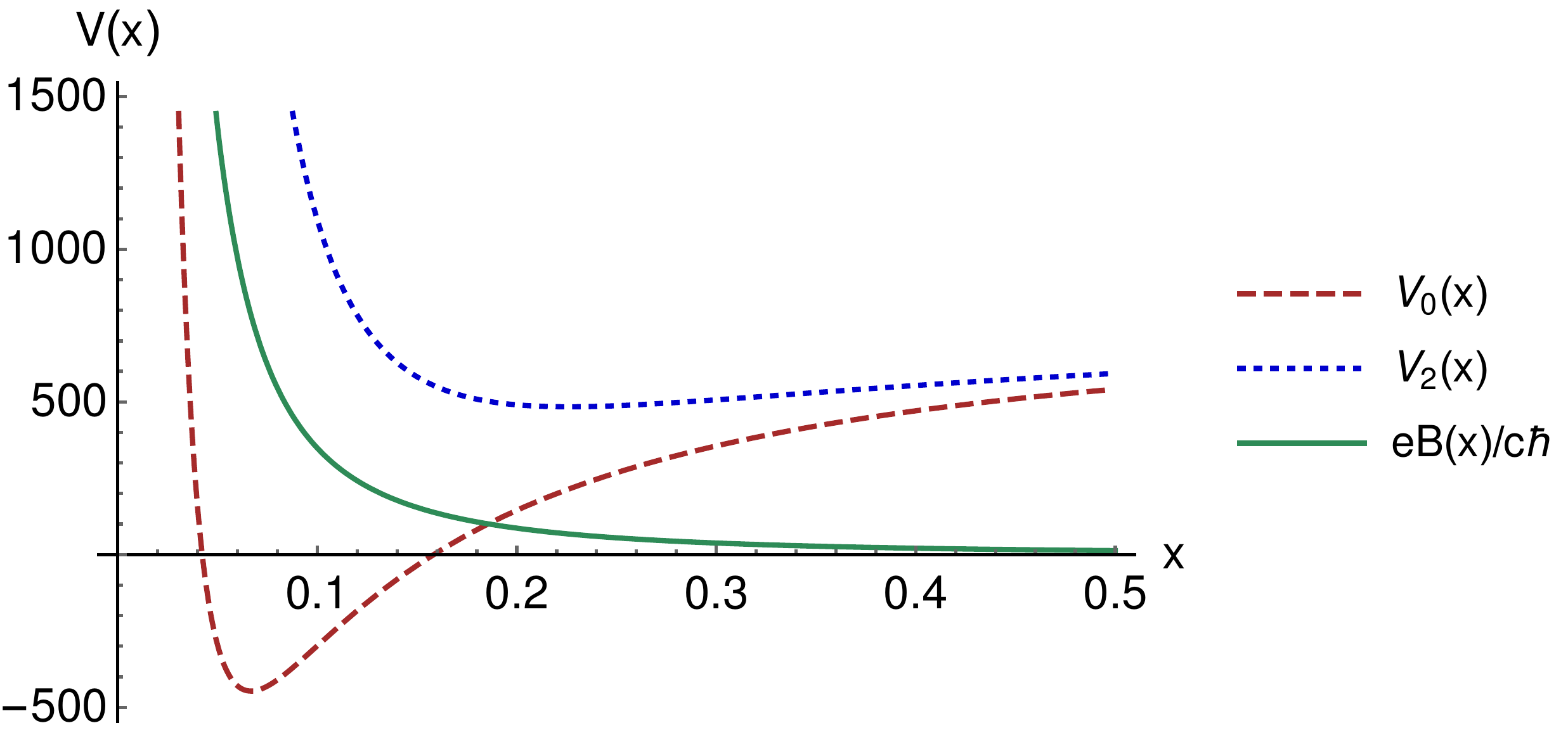}
\includegraphics[width=8.2cm, height=5.5cm]{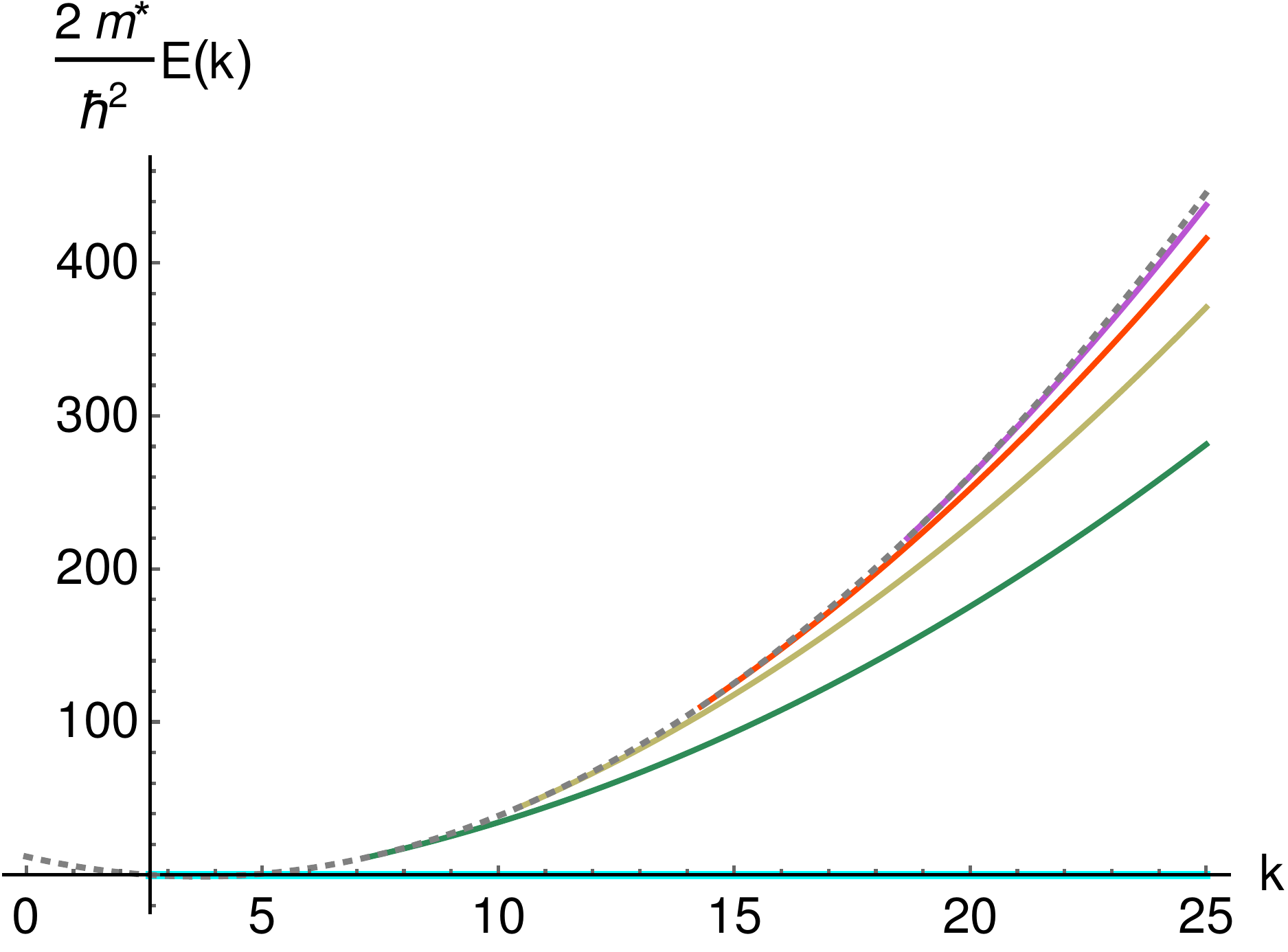}
\caption{Plot of the potentials and the hyperbolic singular magnetic field (left); some eigenenergies $E_n$ as functions of $k$ with ${D}=3$, $k=105/4$ and $\alpha =1$ (right).}
\label{Imagen3_5_1}
\end{figure}

\begin{figure}[ht]
\centering
\includegraphics[width=8.2cm, height=5.5cm]{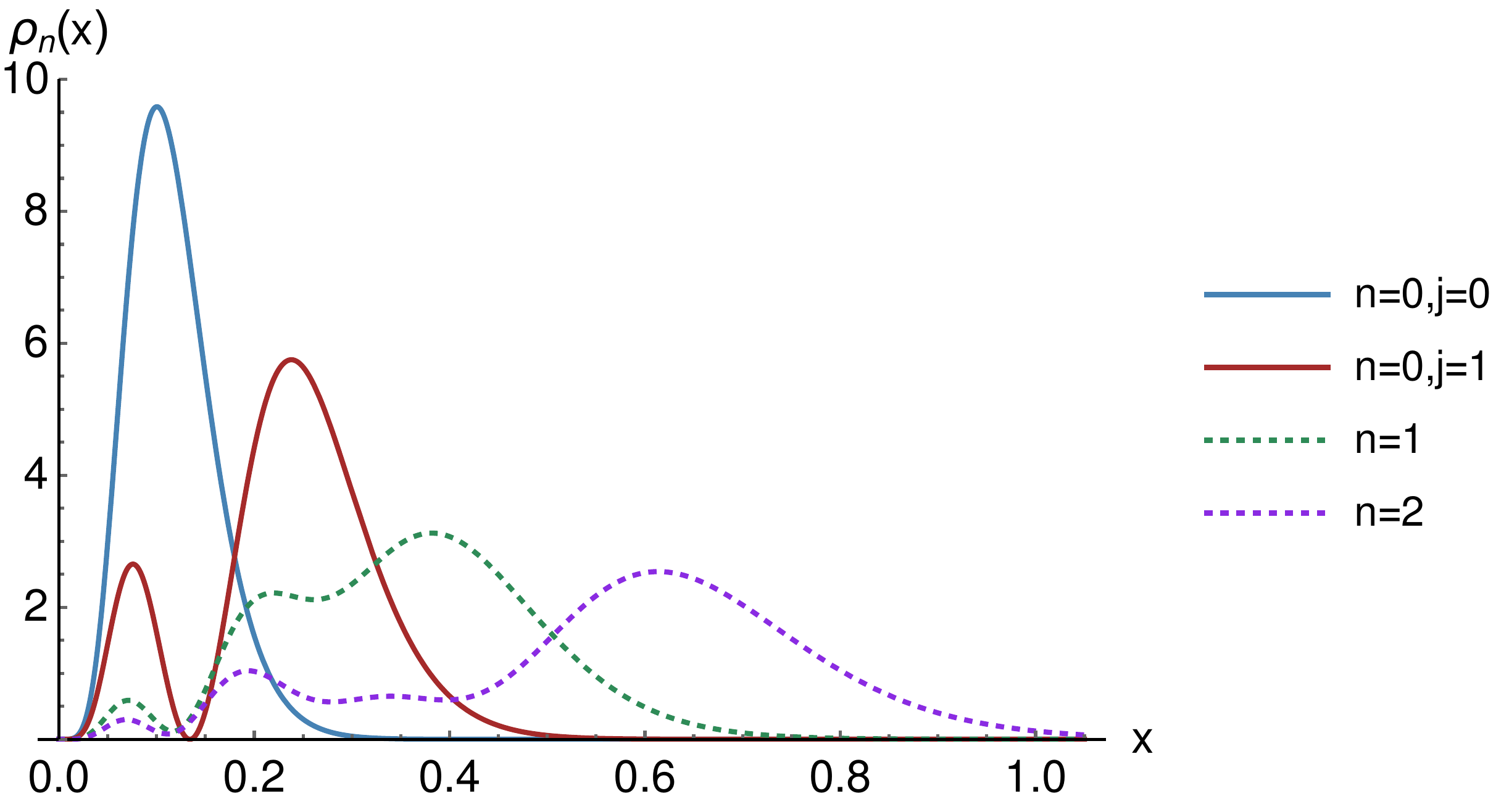}
\includegraphics[width=8.2cm, height=5.5cm]{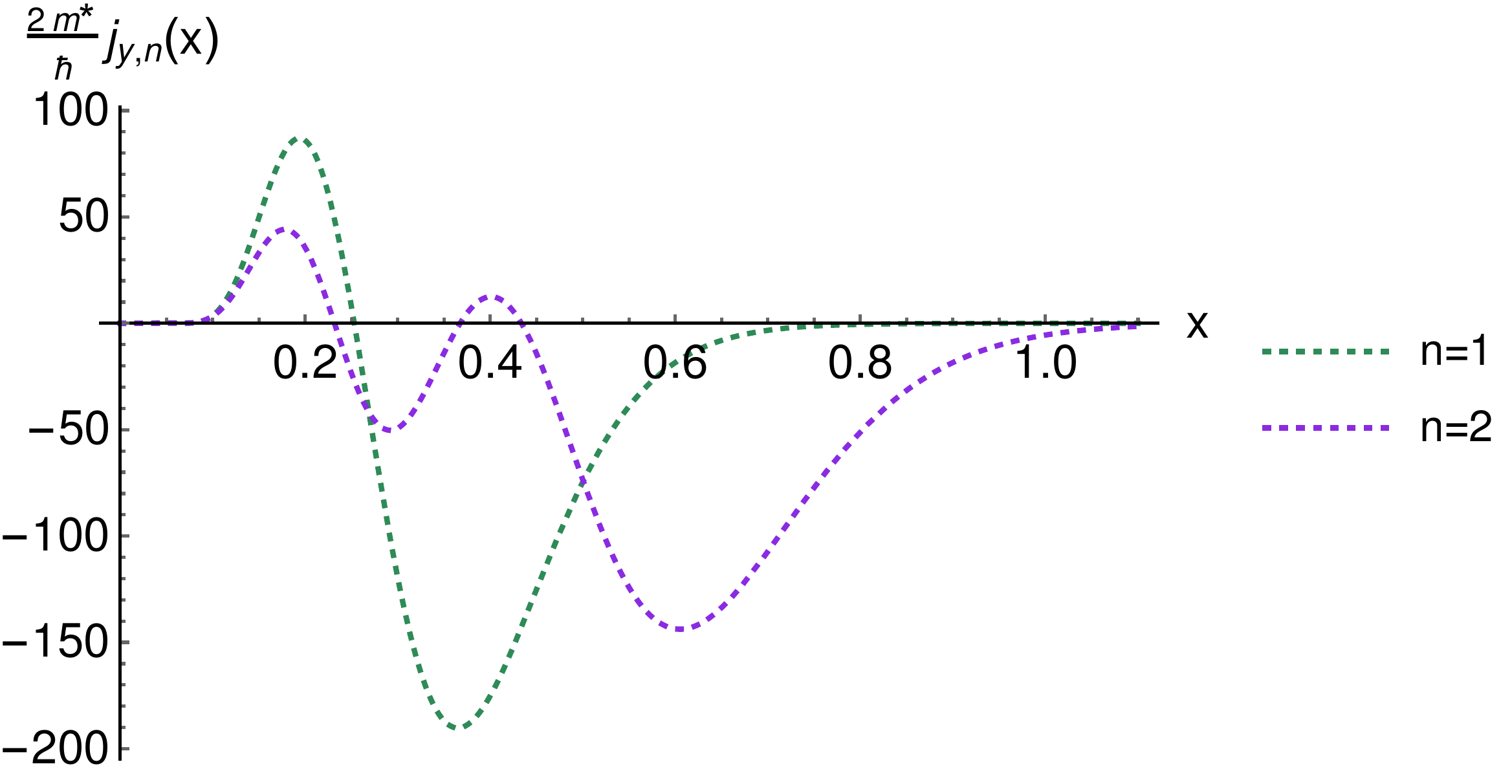}
\caption{Plots of some probability densities (left) and currents (right) for the hyperbolic singular magnetic field with ${D}=3$, $k=105/4$ and $\alpha =1$.}
\label{Imagen3_5_2}
\end{figure}

The eigenvalues of $H$ for electrons in bilayer graphene are now
\begin{equation}
E_{n-1}=\frac{\hbar^{2}}{2m^{*}}\mathcal{E}_{n}^{(0)}\sqrt{1-\gamma_n}, \quad n=1,2,3,\dots,
\label{3.5.8}
\end{equation}
with
\begin{equation}
\gamma_n=
\frac{\kappa^2+{D}^2-({D}+\alpha)^2-\frac{\kappa^2{D}^2}{({D}+\alpha)^2}}{\kappa^2+{D}^2-\left({D}+n\alpha\right)^2-\frac{\kappa^2{D}^2}{\left({D}+n\alpha\right)^2}}.
\label{3.5.9}
\end{equation}
Moreover, an enveloping quadratic polynomial appears, with parameters given by
\begin{equation}
a=4D(D+\alpha)/(2D+\alpha)^{2},\quad b=-2\alpha-4D^{2}/(2D+\alpha),\quad c=D(D+\alpha).
\label{3.5.10}
\end{equation}
Figure \ref{Imagen3_5_1} shows a plot of the potentials and magnetic field (a), while the eigenenergies as functions of $k$ are shown in (b). On the other hand, Figure \ref{Imagen3_5_2} shows a plot of probability densities (a) and currents (b) for some eigenstates of $H$.

\subsection{Case VI: singular magnetic field}
The last case to be considered involves the following magnetic field
\begin{equation}
\vec{B}=\left(0,0,\frac{B_0}{x^{2}}\right),
\label{6.2.1}
\end{equation}
which is generated from the vector potential $\vec{A}=\left(0,-\frac{B_0}{x},0\right)$. According to equation (\ref{2.11}),
the function $\eta$ becomes now
\begin{equation}
\eta(x)=-\frac{(1+2{D})(1+{D}-\kappa x)}{(1+{D})x},
\label{6.2.2}
\end{equation}
where ${D}$ is a dimensionless constant and $\kappa$ has dimension of (length)$^{-1}$, which are given by
\begin{equation}
{D}=\frac{eB_{0}}{\hbar c}-\frac{1}{2},\qquad \kappa=\frac{2(1+{D})}{(1+2{D})}k.
\label{6.2.3}
\end{equation}
If we want to deal with auxiliar exactly solvable shape invariant SUSY partner potentials, we have to choose the factorization energies as $\epsilon_2=0$ and $\epsilon_1=\kappa^2{D}^2\left(\frac{1}{{D}^{2}}-\frac{1}{(1+{D})^{2}}\right)$. Hence:
\begin{align}
V_{0}&=\kappa^2+\frac{{D}({D}-1)}{x^2}-\frac{2\kappa{D}}{x},\label{6.2.4}\\
V_{2}&=\kappa^2+\frac{({D}+2)({D}+1)}{x^2}-\frac{2\kappa{D}}{x},\label{6.2.5}
\end{align}
which are the radial Coulomb potentials with a centrifugal term. Their corresponding eigenenergies are
\begin{align}
\mathcal{E}^{(0)}_{0}&=0,\quad \mathcal{E}^{(0)}_{1}=\kappa^2{D}^2\left(\frac{1}{{D}^{2}}-\frac{1}{(1+{D})^{2}}\right),\nonumber\\
\mathcal{E}^{(0)}_{n}&=\mathcal{E}^{(2)}_{n-2}=\kappa^2{D}^2\left(\frac{1}{{D}^{2}}-\frac{1}{(n+{D})^{2}}\right), \quad n=2,3,\dots 
\label{6.2.6}
\end{align}
The corresponding eigenfunctions are given by  
\begin{align}
\psi_n^{(0)}(\zeta_0)&=\zeta_0^{{D}}e^{\frac{-\zeta_0}{2}}\textrm{L}^{2{D}-1}_n(\zeta_0),\label{6.2.7}\\
\psi_n^{(2)}(\zeta_2)&=\zeta_2^{{D}+2}e^{\frac{-\zeta_2}{2}}\textrm{L}^{2{D}+3}_n(\zeta_2),\label{6.2.8}
\end{align}
where $\zeta_0=\frac{2 \kappa {D}}{n+{D}}x$, $\zeta_2=\frac{2 \kappa {D}}{n+2+{D}}x$, and $\textrm{L}^{a}_{b}$ are the Laguerre polynomials. In order to fulfill the normalizability condition it turns out that $\kappa>0$. 

\begin{figure}[ht]
\centering
\includegraphics[width=8.2cm, height=5.5cm]{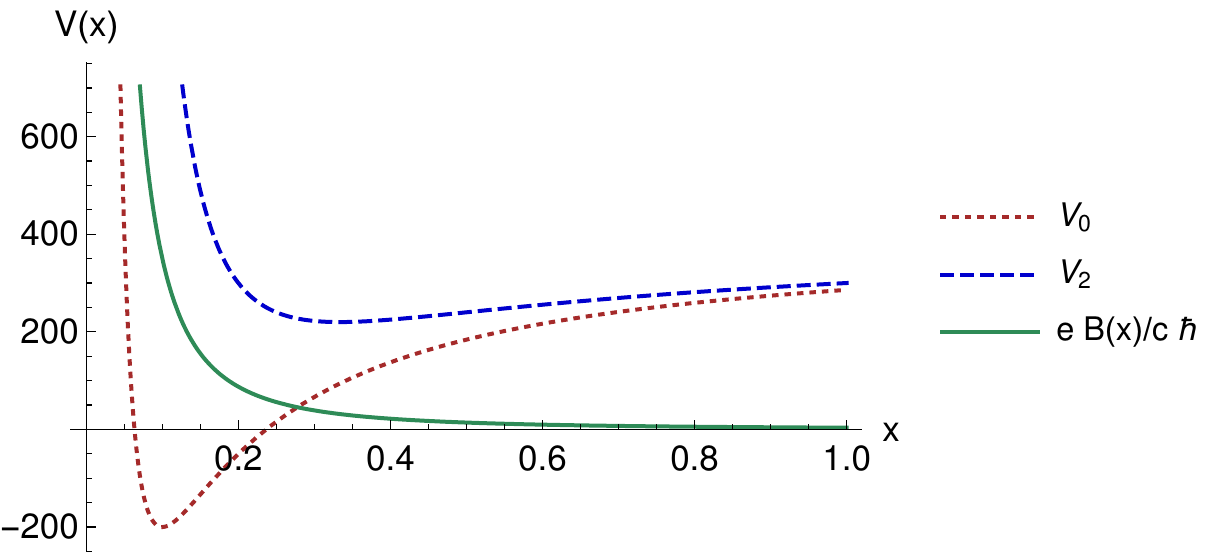}
\includegraphics[width=8.2cm, height=5.5cm]{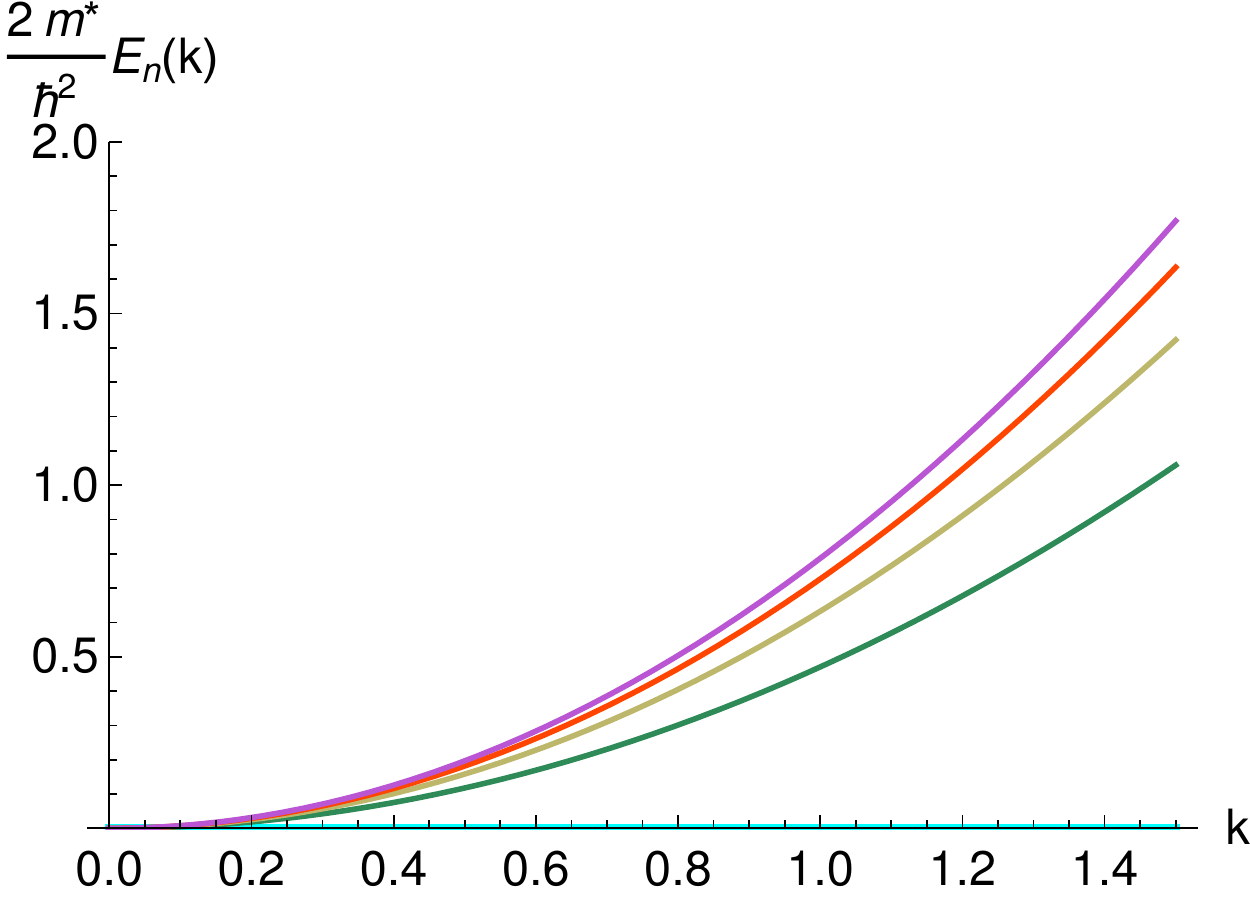}   
\caption{Plot of the SUSY partner potentials $V_0$, $V_2$ and the singular magnetic field (left); some eigenvalues $ E_n$ as functions of $k$ for ${D}=3$ and $k=35/2$ (right).}
\label{Imagen3_6_1}
\end{figure}

\begin{figure}[ht]
\centering
\includegraphics[width=8.2cm, height=5.5cm]{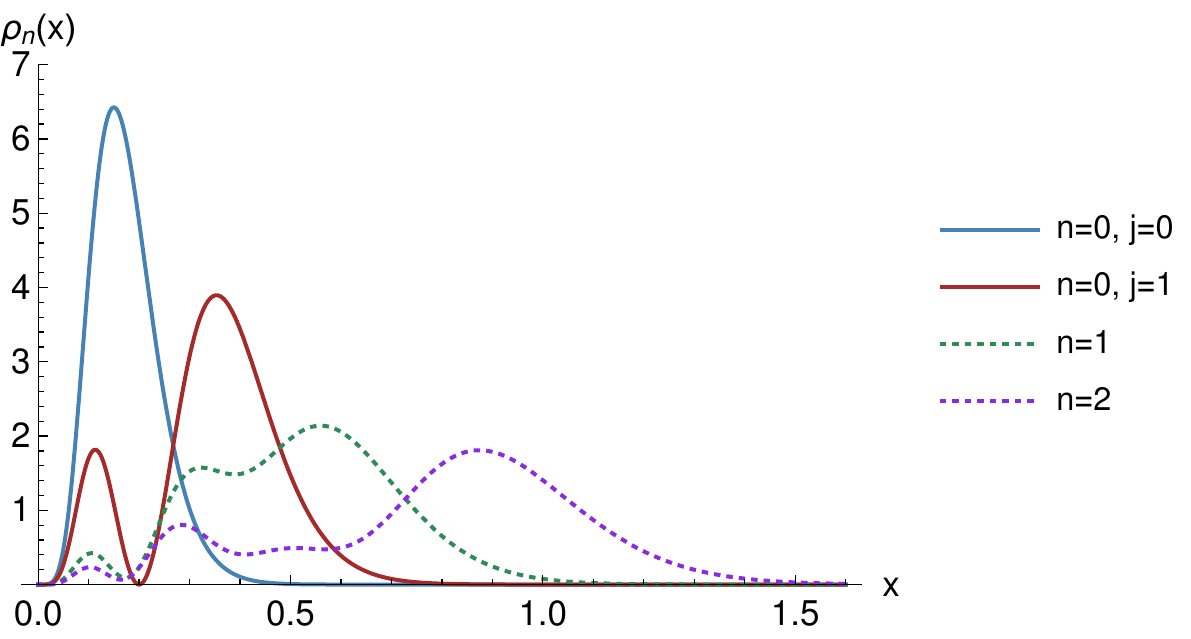}
\includegraphics[width=8.2cm, height=5.5cm]{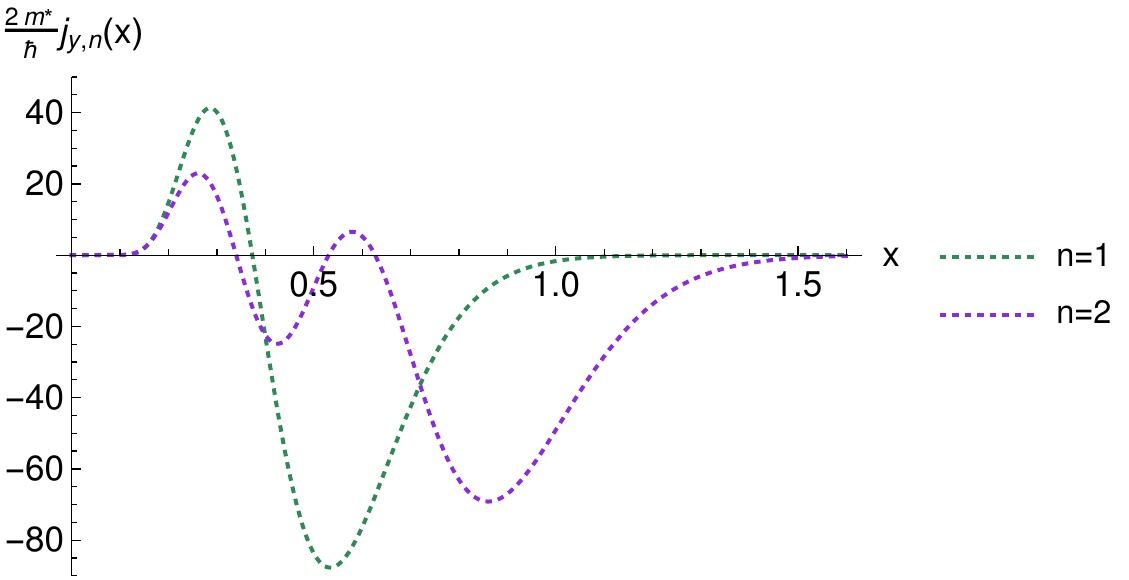}
\caption{Plots of some probability densities (left) and currents (right) for the singular magnetic field with ${D}=3$, $k=35/2$.}
\label{Imagen3_6_2}
\end{figure}

There is a discrete spectrum for $H$, whose energy levels for electrons are
\begin{equation}
E_{n-1}=\frac{\hbar^{2}}{2m^{*}}\mathcal{E}^{(0)}_{n}\sqrt{1-\gamma_n},\quad n=1,2,3,\dots,
\label{6.2.9}
\end{equation}
with
\begin{equation}
\gamma_n = \frac{\kappa^2{D}^2\left(\frac{1}{{D}^{2}}-\frac{1}{(1+{D})^{2}}\right)}{\kappa^2{D}^2\left(\frac{1}{{D}^{2}}-\frac{1}{(n+{D})^{2}}\right)}.
\label{6.2.10}
\end{equation}
As in most of previous cases, the eigenvalues $E_{n}$ depend on the wavenumber $k$. In Figure \ref{Imagen3_6_1} (a) plots of the potentials $V_0$, $V_2$ and the magnetic field are shown, while the eigenvalues $E_n$ as functions of $k$ are drawn in Figure \ref{Imagen3_6_1} (b).
Figure \ref{Imagen3_6_2} sketches the probability densities (a) and currents (b) for fixed values of ${D}$ and $k$.
\newpage
\section{Conclusions}

In this paper the second-order SUSY QM has been implemented successfully to solve the effective Hamiltonian describing the electron motion in bilayer graphene under external magnetic fields orthogonal to the surface inside the tight-binding framework. We have obtained exact analytic expressions for the bound states of the effective Hamiltonian with many different magnetic fields, leading to problems which are translationally invariant along $y$-direction. Moreover, the corresponding auxiliar second-order SUSY partner potentials turn out to be shape invariant, thus exactly solvable. We observe that in most cases the energy eigenvalues have an explicit dependence on the wavenumber $k$, except for the constant magnetic field. Unlike the monolayer graphene, here we obtain a double degenerate ground state energy level. We have to stress also on the existence of cases for which the energy spectrum is discrete and finite, depending on the wavenumber $k$. For the cases where this happens (Cases II, IV and V), we have built an enveloping quadratic polynomial, which touches the end point energies where the bound states transform into scattering states. Due to this quadratic dependence on $k$, we can show that the group velocity in $y$-direction is not a constant, and that the only non-zero component of the effective mass tensor will be constant, with its value just depending on the parameters $D$ and $\alpha$. In one of these cases (Case IV) the constants $a$, $b$ do not depend on $D$, which is linear in the field amplitude $B_0$, but in all three cases presented here the effective mass keeps constant and the group velocity depends linearly on the value of $k$, regardless the values of the magnetic field parameters. However, there are two cases (Cases II and V) in which there is an explicit dependence on the parameters of the field, so the limits $D\rightarrow0$ and $D\gg0$ are worth of some study. For the first limit ($D\rightarrow0$) it is observed that the group velocity becomes a constant, which implies that the effective mass tends to infinity. For the second limit ($D\gg0$) it is seen that the effective mass remains constant, but not the group velocity.  We have to emphasize that this analysis is valid only for the gauge (Landau) that has been chosen, which is given by $(0,A_y(x),0)$. A different analysis should be carried out if the gauge chosen would depend on the $y$ coordinate \cite{dnn19}, since the proposal of equation~(\ref{2.7}) would lose sense due to the lack of translational invariance along $y-$direction.

\section*{Acknowledgments}
JDGM (number 487715) and DOC especially thank Conacyt for the economic support through the PhD scholarships.
\section*{References}
\bibliography{biblio}
\bibliographystyle{iopart-num}
\end{document}